Title: AN INTERFACE-TRACKING TECHIQUE FOR MULTIPHASE FLOW WITH SOLUBLE SURFACTANT

Article Type: Regular Article

Corresponding Author: Dr. Seungwon Shin, PhD

Corresponding Author's Institution: Hongik University

First Author: Seungwon Shin, PhD

Order of Authors: Seungwon Shin, PhD; Jalel Chergui, Ph.D; Damir Juric, Ph.D; Lyes Kahouadji, Ph.D; Omar K Matar, Ph.D; Richard V Craster, Ph.D

Abstract: We adapt and extend a formulation for soluble surfactant transport in multiphase flows recently presented by Muradoglu & Tryggvason (JCP 274 (2014) 737-757) to the context of the Level Contour Reconstruction Method (Shin et al. IJNMF 60 (2009) 753-778) which is a hybrid method that combines the advantages of the Front-tracking and Level Set methods. Particularly close attention is paid to the formulation and numerical implementation of the surface gradients of surfactant concentration and surface tension. Various benchmark tests are performed to demonstrate the accuracy of different elements of the algorithm. To verify surfactant mass conservation, values for surfactant diffusion along the interface are compared with the exact solution for the problem of uniform expansion of a sphere. The numerical implementation of the discontinuous boundary condition for the source term in the bulk concentration is compared with the approximate solution. Surface tension forces are tested for Marangoni drop translation. Our numerical results for drop deformation in simple shear are compared with experiments and results from previous simulations. All benchmarking tests compare well with existing data thus providing confidence that our adapted LCRM formulation for surfactant advection and diffusion is accurate and effective in three-dimensional multiphase flows. We also demonstrate that this approach applies easily to massively parallel simulations.

Suggested Reviewers: Metin Muradoglu
Ph.D, Department of Mechanical Engineering, Koc University
mmuradoglu@ku.edu.tr
Expert in surfactant modeling for front tracking method

Alexandre M Roma Ph.D
Instituto de Matemática e Estatística, Universidade de São Paulo
roma@ime.usp.br
Expert in surfactant modeling with front tracking method

Dieter Bothe Ph.D

Department of Mathematics, University of Darmstadt
bothe@mma.tu-darmstadt.de
Expert in numerical simulation of two phase flows and surfactant

# Highlights (Significance and Novelty of the paper)

- Extension of the LCRM Front-tracking method (Shin et al. IJNMF 60 (2009) 753-778) to two-phase flows with surfactant with close attention paid to the formulation and numerical implementation of surface gradients of surfactant concentration and surface tension.

- Following the theoretical framework in Muradoglu & Tryggvason (JCP 274 (2014) 737-757), surfactant transport is solved both on the interface and in the bulk liquid phase.

- Benchmark tests demonstrate accuracy for different elements of the numerical implementation: mass conservation, surface advection and diffusion, bulk transport and Marangoni stresses.

- Large scale parallel calculations of two-phase annular film flow in the counter-current flow regime both with and without the presence of surfactant and both with and without surfactant solubility exhibiting features expected in such a complex flow: wave and ligament formation, droplet detachment and entrainment and flooding.

# AN INTERFACE-TRACKING TECHIQUE FOR MULTIPHASE FLOW WITH SOLUBLE SURFACTANT

Seungwon Shin[1,*], Jalel Chergui[2], Damir Juric[2]

Lyes Kahouadji[3], Omar K. Matar[3] and Richard V. Craster[4]

[1] Department of Mechanical and System Design Engineering, Hongik University, Seoul, 121-791 Korea

[2] Laboratoire d'Informatique pour la Mécanique et les Sciences de l'Ingénieur (LIMSI),

Centre National de la Recherche Scientifique (CNRS),

CNRS-UPR 3251, Bât. 508, Rue John von Neumann, Campus Universitaire d'Orsay, 91405 Orsay, France

[3] Department of Chemical Engineering, Imperial College London,

South Kensington Campus, London SW7 2AZ, UK

[4] Department of Mathematics, Imperial College London,

South Kensington Campus, London SW7 2AZ, UK

[*]Corresponding author :

    Seungwon Shin, Ph.D

    Associate Professor

    Department of Mechanical and System Design Engineering

    Hongik University

    Sangsu-dong, 72-1, Mapo-gu

    Seoul, 121-791, Korea

    Phone: 82-2-320-3038

    FAX: 82-2-322-7003

    E-Mail: sshin@hongik.ac.kr

# ABSTRACT

We adapt and extend a formulation for soluble surfactant transport in multiphase flows recently presented by Muradoglu & Tryggvason (JCP 274 (2014) 737-757) to the context of the Level Contour Reconstruction Method (Shin et al. IJNMF 60 (2009) 753-778) which is a hybrid method that combines the advantages of the Front-tracking and Level Set methods. Particularly close attention is paid to the formulation and numerical implementation of the surface gradients of surfactant concentration and surface tension. Various benchmark tests are performed to demonstrate the accuracy of different elements of the algorithm. To verify surfactant mass conservation, values for surfactant diffusion along the interface are compared with the exact solution for the problem of uniform expansion of a sphere. The numerical implementation of the discontinuous boundary condition for the source term in the bulk concentration is compared with the approximate solution. Surface tension forces are tested for Marangoni drop translation. Our numerical results for drop deformation in simple shear are compared with experiments and results from previous simulations. All benchmarking tests compare well with existing data thus providing confidence that our adapted LCRM formulation for surfactant advection and diffusion is accurate and effective in three-dimensional multiphase flows. We also demonstrate that this approach applies easily to massively parallel simulations.

# 1. INTRODUCTION

The dynamics of the interface which separate two distinct fluids is heavily dependent on the interfacial surface tension. For a constant surface tension coefficient, the component of the surface tension force normal to the interface exerts an inward "pull". Any imbalance between this normal inward tension and normal pressure gradient across the interface will result in interface movement. Due to the complexities associated with general multiphase flows where viscous, inertial and possibly other (electromagnetic) forces play a role, numerical analysis has provided a vital tool for the study of the detailed physical processes associated with interface motion in such flows.

Most of the numerical investigations in surface tension-driven flows have focused on the accurate tracking, or capturing, of the moving interface for the specific case where the surface tension coefficient is constant. Popular approaches are Front-tracking [1], Volume of Fluid (VOF) [2], Level Set [3], Lattice Boltzmann [4], and Phase Field [5] methods. The latter four methods, falling in the category of front-capturing, require an additional scalar advection equation to track the interface motion. On the other hand, the Front-Tracking method employs a separately tracked Lagrangian grid of interface elements, which provides a precise location of the interface front and thereby a robust and accurate curvature (derivative) free approach to the calculation of the surface tension force. Recently, hybrids of the above methods have appeared, which attempt to retain the advantages of one approach while avoiding the inconvenient aspects of the other; Tryggvason et al. [6] provide an excellent overview and introduction to recent work in this area. Our Level Contour Reconstruction Method (LCRM) [7, 8] is one such hybrid method which retains the accuracy of Front-tracking of a separate Lagrangian interface while also retaining the ease with which topological coalescence and rupture of interfaces are handled by the Level Set method (while completely avoiding Level Set's well known mass conservation difficulties).

By adding surfactant to the interface, the surface tension coefficient is no longer uniform along the interface. The surface active agents (surfactants) are usually amphiphilic organic compounds, which adsorb onto, or desorb from, the interface. They also typically consist of hydrophilic heads and hydrophobic tails and thus tend to accumulate at the interface separating two phases. Surfactants act to locally decrease the surface tension coefficient and thereby change the interfacial dynamics significantly. A non-uniform surface tension coefficient along the interface will generate tangential forces, i.e. the Marangoni effect. Many microfluidic devices use these forces to control drop dynamics by adding surfactant to the interface thus changing the surfactant concentration distribution along the interface [9, 10]. These Marangoni stresses pose additional challenges to formulating and implementing an accurate numerical interface method, which allows surfactants to adsorb, or desorb, onto or from the interface.

Various attempts have been made to solve numerically the surfactant transport equation. Renardy *et al.* [11] used the VOF method (SURFER++) to investigate deformation of a drop under simple shear flow. A linear equation of state for the continuous surface stress was formulated to include surfactant effects. In their simulations, only the bulk concentration equation was considered and the effect on the surface tension coefficient was introduced through a reduction factor model. Drop deformation in shear stress was analyzed under various input conditions. James and Lowengrub [12] developed surfactant equations in axisymmetric geometry. The interface motion was captured by the VOF method and the surface tension force including surfactant effects were implemented in the momentum equation via the continuum surface force method [13]. A convection-diffusion equation for surfactant on the interface was solved to simulate drop deformation in extensional flow. They used a source term to account for interfacial stretching and a Lagrangian description of the surfactant on the interface to evaluate the concentration gradient. Detailed convergence tests and validations were performed to show the accuracy of their method but due to axi-symmetry they were limited to performing simpler drop extension and retraction tests.

Feigl *et al.* [14] investigated droplet deformation in simple shear flow using coupled boundary integral, interface-tracking, and finite element methods. The boundary integral method was used to solve for the interfacial velocity and the finite element method for surfactant concentration. The relation between surfactant concentration and surface tension coefficient was modeled using a linear equation of state. Their algorithm was validated by comparing with experimental results for drop deformation with different viscosity ratios. They found that the surfactant tends to make the drops align along the flow direction. Even though they considered three-dimensional (3D) geometries, coordinate transformations were necessary to represent the surfactant governing equation in two-dimensional surface coordinates embedded in three-dimensional space.

Teigen *et al*. [15] presented a diffuse-interface method to solve the behavior of a drop in simple shear flow for soluble surfactants. A block structured adaptive grid was used for additional grid refinement. They used a non-linear form of the equation of state to represent the functional relationship between the surface tension coefficient and surfactant concentration. Several benchmark tests were performed to verify accuracy by simulating the oscillation of a capillary wave and a rising drop in a linear surfactant gradient (i.e. Marangoni migration). Drop stretching in a linear flow was also tested for small capillary numbers and compared with the theoretical solution. Furthermore, they conducted a detailed investigation of changing desorption and adsorption coefficients on drop stretching characteristics. Most of their simulations were performed in two-dimensions while a few 3D cases were tested. They chose to represent the interface implicitly via a phase-field model which required the solution of a fourth-order in space Cahn-Hilliard system. The simulations were CPU time-intensive for a full 3D case even with adaptive grid refinement.

Chen and Lai [16] used an immersed boundary formulation to solve the surface-bulk concentration equation. They focused on developing a conservative scheme which can preserve the total mass of the surfactant in a discrete sense. The interface was represented by Lagrangian elements and a Dirac delta function was used in the immersed boundary technique. Both desorption and adsorption at the interface were considered when solving the surface concentration equation. Various validation tests were performed to show the accuracy of their method including drop deformation under shear flow. However their simulations were limited to two-dimensions.

Muradoglu and Tryggvason [17] developed a finite-difference numerical procedure based on the Front-Tracking method to simulate soluble surfactants in three-dimensions. They extended their previous axisymmetric work [18] to fully-3D simulations of the diffusion process on the interface and source boundary conditions for the bulk phase. The surface gradient on the Lagrangian interface was computed at the nodes of triangular elements by summing contributions from neighboring elements. Thus, it is important to note that their method requires the connectivity information of classical-Front-tracking. A special treatment for the source term in the bulk phase concentration equation was also devised to allow adsorption or desorption from the surface to one side of the bulk phase. The source term computed on the interface was distributed over the absorption layer placed outside of the drop region in a conservative manner. Several discretization schemes for the convective term were tested and the up-winded WENO_Z scheme was chosen due to its smoothness and simplicity. Various benchmarking tests were performed to identify the accuracy of their numerical procedure including surface expansion, surfactant diffusion on the interface, Marangoni migration, bulk surfactant convection and diffusion, and buoyancy-driven bubble rise. They then applied their approach to bubble motion in a pressure-driven channel flow and found that surfactant on the interface can affect lateral bubble motion significantly by counteracting the shear induced lift force with surfactant induced Marangoni stress.

De Jesus *et al*. [19] developed a full, 3D numerical procedure for the advection and diffusion equations for insoluble surfactant based on a Front-Tracking technique. Adaptive mesh refinement was used in a finite volume discretization of the governing equations. A slightly different formulation for the surfactant diffusion was considered for interface projection and mesh optimization. Mesh optimization was performed to prevent degeneration of the Lagrangian interface elements due to clustering by forcing a slight slide in the tangential direction. Special treatment is necessary to keep each circumcenter of the element inside its triangle. Surface convolution was tested for surfactant mass conservation performance and drop deformations in simple shear flow were compared with experimental and other simulation results. The results matched the existing solution fairly well. Even though adaptive mesh refinement makes it possible to resolve more detailed features near the interface and save computational time and memory, no topological changes are allowed. Various attempts to use different interface-tracking procedures to solve the surfactant transport equation have been devised. Xu *et al*. [20] proposed a level-set continuum surface force method for full 3D two-phase flow with insoluble surfactant. A moving particle semi-implicit method was also used to solve multiphase flow with soluble surfactant [21].

In these recent studies, full 3D simulations have been attempted mostly for cases with rather simple dynamics (simple shear stress in drop deformation or bubble rise with surfactant). This is partly due to the challenges in the treatment of surfactant transport on the interface. Surfactant will be transported along the interface from high to lower concentrations and the interface can possibly undergo complex topological changes thus diffusion on arbitrary surfaces can be handled more naturally in a Lagrangian way. In methods using an Eulerian approach for surfactant transport the formulation tends to become complicated and implementation of a general procedure is a cumbersome task. In our approach to interface-tracking, we use a hybrid type method (LCRM) combining the advantages of Front-Tracking and Level Set methods. In this paper, we present our extension and adaptation of the formulation for soluble surfactant transport in multiphase flows recently presented by Muradoglu & Tryggvason [17]. We have implemented this approach in our code, BLUE [22], a massively-parallel multiphase flow solver, and we discuss newly-developed solution techniques for surfactant transport suitable for distributed processing. Particularly close attention was paid to the formulation and numerical implementation of the surface gradients of surfactant concentration.

The rest of this paper is organized as follows: we briefly discuss the general interface-tracking method (LCRM) in the next section. Then we present a detailed description of the surfactant transport equations and solution procedure for surface diffusion and interface force model for varying surface tension coefficient. The discontinuous boundary condition implementation for the source term in the bulk surfactant equation will also be discussed. Finally, we present various benchmark tests to demonstrate the accuracy of different elements of the algorithm.

# 2. NUMERICAL FORMULATION

2.1 Interface-tracking method

We use the Level Contour Reconstruction method (LCRM) to track the moving interface [18]. LCRM is a hybrid interface method combining advantages of classical Front-tracking and Level Set front-capturing. In Front-tracking, besides the usual Eulerian grid for the resolution of field equations, an additional moving and deforming two-dimensional (2D) Lagrangian grid composed of interface elements (triangles) is tracked in 3D space. The Lagrangian grid makes it possible to advect the interface and associated quantities such as discontinuous property fields (density, viscosity) accurately without undesired numerical smearing over time. However, classic Front-tracking, as used in [17], requires special routines to keep track of element connectivity information as well as for element size regularization as the interface stretches and deforms. For 3D simulations, these routines become quite complicated particularly when handling topological changes of the interface such as rupture and coalescence. In addition, these routines are difficult to parallelize since connectivity information must be communicated to all processors. On the other hand, front-capturing type methods locate the interface position implicitly by numerically advecting an additional Eulerian scalar field, i.e. a distance function for Level Sets or the volume fraction for the Volume of Fluid method. Since this additional scalar field uses the same Eulerian grid structure as velocity and pressure variables, parallelization does not pose additional challenges, and topological changes are handled implicitly. However, the main drawbacks of front-capturing methods are in accurate surface tension force calculation, or maintaining mass conservation (Level Sets) without additional remedies.

Our LCRM is founded on Front-tracking since it tracks an additional Lagrangian interface. It avoids the problem of complicated element connectivity since the Lagrangian interface is entirely reconstructed from time to time by using information from a Level Set type distance function field. Since tracking provides the precise location of the interface elements at all times, it is straightforward to construct the distance function field when needed for the interface reconstruction step. Thus the LCRM also avoids the drawbacks of the Level Set method since it never advects a scalar distance function field. Fig.1 shows the basic concept of the Level Contour Reconstruction Method in 3D. With a given set of Lagrangian interface elements, we can directly generate a distance function field (the detailed procedure can be found in [8]). Reconstruction is performed on the same Eulerian grid structure as the velocity and pressure. In order to implicitly connect triangular elements which inhabit neighboring Eulerian cells we subdivide these cells into tetrahedra (Fig. 1(b)). The main idea of this cell-by-cell reconstruction is to simply draw the zero isocontour surface of the distance function field and identify the intersection of this surface with the tetrahedra. This tetra-marching procedure guarantees that at most one unique isocontour surface can be identified in each reconstruction cell; the marching cube method on the other hand could result in ambiguity. The user can control the frequency of interface reconstruction but typically it is performed about once every 25 time steps. The purpose of the reconstruction is to regularize the interface elements from time to time to avoid excessively large elements and to avoid dispersion and coagulation of elements due to interface deformation. In addition topological changes to the interface are automatically handled since newly reconstructed elements will take on the topological characteristics of the distance function. A high order reconstruction technique [23] improves the precision of element positioning during reconstruction. The computation of surface tension forces was also improved [24] to eliminate spurious numerical parasitic currents. The accurate calculation of surface tension forces in the presence of surfactant will be discussed in more detail in subsequent sections as it applies to the surface forces for varying surface tension.

Recently, we have developed a general purpose 3D massively parallel multiphase flow code [22]. This code is particularly suited to direct simulation of incompressible surface tension-driven flows. A parallel hybrid Multigrid/GMRES algorithm efficiently solves the pressure Poisson equation even at very high density ratio of $O(10^4)$. Parallelization of the LCRM with MPI and domain-decomposition is straightforward since all interface operations are local to an element and its local region of grid cells and this characteristic feature of the LCRM is inherited by each subdomain. Various modules of BLUE are dedicated to a wide variety of multiphase scenarios and the code has been rigorously tested on a suite of multiphase benchmark problems as well as academic research and problems in the chemical and pharmaceutical industries.

## 2.2 Governing equations

The governing equations of mass and momentum conservation for an incompressible multiphase flow can be described in a single field form as follows:

$$\nabla \cdot \mathbf{u} = 0 \tag{1}$$

$$\rho\left(\frac{\partial \mathbf{u}}{\partial t} + \mathbf{u} \cdot \nabla \mathbf{u}\right) = -\nabla P + \rho \mathbf{g} + \nabla \cdot \mu(\nabla \mathbf{u} + \nabla \mathbf{u}^\mathrm{T}) + \mathbf{F} \tag{2}$$

where $\mathbf{u}$ is the velocity, $P$, the pressure, $\mathbf{g}$, the gravitational acceleration, and $\mathbf{F}$ represents the surface tension force (section 2.4 presents a detailed formulation of $\mathbf{F}$ for variable surface tension coefficient). For two phases, the Heaviside function, $I$, is zero in one phase and one in the other and is used to define material properties such as density and viscosity:

$$\rho = \rho_1 + (\rho_2 - \rho_1)I \tag{3}$$

$$\mu = \mu_1 + (\mu_2 - \mu_1)I \tag{4}$$

Here, subscripts denote the individual phases. The Heaviside function is constructed from the distance function field which is computed directly from location of the Lagrangian interface [8].

The surfactant concentration on the interface, $\Gamma$, is governed by the following conservation equation [17]:

$$\frac{\partial \Gamma}{\partial t} + \nabla_s \cdot (\Gamma \mathbf{U}_s) = D_s \nabla_s^2 \Gamma + \dot{S}_\Gamma \tag{5}$$

The left hand side of Eq. (5) describes transient and convective transport of the surfactant concentration on the interface. $\mathbf{U}_s$ represents the tangential velocity vector on the interface and $\nabla_s$ represents the surface gradient operator ($\nabla$-**n**(**n**·$\nabla$)), **n** being the vector normal to the interface. The first term on the right-hand-side of Eq. (5) accounts for diffusion of the surfactant along the interface where $D_s$ is the diffusion coefficient. $\dot{S}_\Gamma$ is a source term due to adsorption and desorption between the bulk flow and interface and is described by:

$$\dot{S}_\Gamma = k_a C_s \left( \Gamma_\infty - \Gamma \right) - k_d \Gamma \tag{6}$$

Here, $k_a$ and $k_d$ are the adsorption and desorption coefficients, respectively, $C_s$ is the concentration of surfactant in the bulk fluid immediately adjacent to the interface, and $\Gamma_\infty$ is the maximum packing interfacial concentration.

The governing transport equation for bulk surfactant concentration $C$ is:

$$\frac{\partial C}{\partial t} + \mathbf{u} \cdot \nabla C = \nabla \cdot \left( D_c \nabla C \right) \tag{7}$$

Here, $D_c$ represents the bulk diffusion coefficient. We assume the surfactant to be soluble in just phase 2 and hence we allow bulk diffusion only in one phase and thus we define $D_c$ as:

$$D_c = D_{c2} I(\mathbf{x}, t) \tag{8}$$

where $D_{c2}$ is the bulk diffusion coefficient in phase 2. The source term in Eqs. (5) and (6) can be related to the bulk concentration by:

$$\mathbf{n} \cdot \nabla C \big|_f = -\dot{S}_\Gamma / D_{c2} = \frac{\partial C}{\partial \mathbf{n}} \bigg|_f \tag{9}$$

where, $f$ represents the interface location. This condition is enforced by a sharp boundary method [25]

and is described in detail in Section 2.5.

The interfacial elements are advected in Lagrangian fashion by integrating

$$\frac{d\mathbf{x}_f}{dt} = \mathbf{V} \qquad (10)$$

with a second-order Runge-Kutta method where the interface velocity, **V**, is interpolated from the Eulerian velocity. The well-known projection method on a staggered MAC mesh is used to solve for fluid velocity and pressure. A second-order ENO scheme is used for convective terms. A more detailed description of the procedure for the solution of the momentum equation can be found in [7-8, 22-25].

2.3 Surfactant conservation equation

To solve the surfactant conservation equation on the evolving interface we essentially follow the derivation of Muradoglu and Tryggvason [17] except for the surface diffusion term in Eq. (5). We prefer to maintain the LCRM's ease of parallelization and thus implement this term in a different manner. Using the Leibniz formula, the transient and convective transport terms on the left of Eq. (5) can be discretized as:

$$\int_{A_e} \left[ \frac{\partial \Gamma}{\partial t} + \nabla_s \cdot (\Gamma \mathbf{U}_s) \right] dA_e \approx \frac{(\Gamma A_e)^{n+1} - (\Gamma A_e)^n}{\Delta t} \qquad (11)$$

Here, $A_e$ represents the area of the triangular element $e$ on the Lagrangian interface as described in Fig. 2(a); $n$ and $n+1$ denote the time step level. An element is composed of three edge lines $\overline{①②}$, $\overline{②③}$, and $\overline{③①}$ forming a triangle. The right side of Eq. (5) can be converted into discrete form similarly as:

$$\int_{A_e} \left( D_s \nabla_s^2 \Gamma + \dot{S}_\Gamma \right) dA = \delta \Gamma_{D_e}^n + \left( \dot{S}_\Gamma A_e \right)^n \tag{12}$$

Thus the discretized form of the surfactant conservation equation can be easily integrated over time except for the first term on the right hand side of Eq. (12) which represents diffusion of the surfactant along the interface. To compute this diffusion term, a complete description of the Lagrangian interface is necessary since the surface gradient of the surfactant concentration is heavily dependent on the geometry of the interface. Most previous work has used the Front-tracking concept to compute this surface gradient [12, 14, 16-19]. In some cases, an Eulerian function such as a distance function can be used but the formulation becomes extremely complex [15, 20]. The Front-tracking method has a unique advantage for the computation of surface gradient since interface connectivity can be effectively used to calculate the surface gradient term. On the other hand, this connectivity has its own drawback since it becomes cumbersome in 3D especially when dealing with topology change of the interface and also in parallel implementation.

To avoid these complexities we prefer the hybrid LCRM as described in the previous section. To fully use the Eulerian component of the LCRM scheme, the diffusion of the surfactant along the interface should be reformulated to be compatible with an Eulerian structure. We describe the geometrical information of the target interface element in Fig. 2(a). Triangular interface element $e$ has area $A_e$ with interface normal $\mathbf{n}_f$ at its center. The surface surfactant concentration $\Gamma_f$ is stored at the center of the element. Since a Lagrangian element is defined by vertices ①, ②, and ③, the area and normal of the element can be easily computed:

$$A_e = \frac{1}{2} \left| \mathbf{x}_{①②} \times \mathbf{x}_{①③} \right| \tag{13}$$

$$\mathbf{n}_f = \frac{\mathbf{x}_{①②} \times \mathbf{x}_{①③}}{\left| \mathbf{x}_{①②} \times \mathbf{x}_{①③} \right|} \tag{14}$$

The surface gradient ($\nabla_s$) is computed at the middle of element edge points. $\Gamma_{12}$, $\Gamma_{23}$, and $\Gamma_{31}$ are the interpolated values at the centers of the edges at $\overline{①②}$, $\overline{②③}$, and $\overline{③①}$, respectively. The surface normal at these center edge locations can be computed using the distance function field as follows:

$$\mathbf{n}(\mathbf{x}) = \frac{\nabla \phi(\mathbf{x})}{|\nabla \phi(\mathbf{x})|} \tag{15}$$

In the LCRM formulation, the distance function field ($\phi$) is obtained directly from the tracked interface elements by finding the minimum distance value to the interface elements. The edge normals at the centers of the edges, $\mathbf{n}_{12}$, $\mathbf{n}_{23}$, and $\mathbf{n}_{31}$, can be used to compute the binormal vectors $\mathbf{p}_{12}$, $\mathbf{p}_{23}$, and $\mathbf{p}_{31}$ as described in Fig. 2(a) by the cross product ($\mathbf{n} \times \mathbf{t}$) at the given locations. Slightly different binormal vectors denoted by $\mathbf{p}'_{12}$, $\mathbf{p}'_{23}$, and $\mathbf{p}'_{31}$ in Fig. 2(a) are computed using $\mathbf{n}_f$ and $\mathbf{t}$ and these will later be used to compute the interfacial force.

Following the formulation of Muradoglu and Tryggvason [17], the surface gradient term can be evaluated as a line integral along the edges of the element:

$$\delta \Gamma_{D_e}^n = D_s \iint_{A_e} \nabla_s^2 \Gamma dA = D_s \oint_c \nabla_s \Gamma \cdot \mathbf{p} ds \tag{16}$$

This line integral along the edges of a particular triangular element can be computed as:

$$\begin{aligned} \delta \Gamma_{D_e}^n &= D_s \sum_{k=1}^{3} (\nabla_s \Gamma)_k \cdot \mathbf{p}_k \Delta s_k \\ &= D_s \left\{ (\nabla_s \Gamma)_{12} \cdot \mathbf{p}_{12} \Delta s_{12} + (\nabla_s \Gamma)_{23} \cdot \mathbf{p}_{23} \Delta s_{23} + (\nabla_s \Gamma)_{31} \cdot \mathbf{p}_{31} \Delta s_{31} \right\} \\ &= D_s \left\{ (\nabla_s \Gamma)_{p12} \Delta s_{12} + (\nabla_s \Gamma)_{p23} \Delta s_{23} + (\nabla_s \Gamma)_{p31} \Delta s_{31} \right\} \end{aligned} \tag{17}$$

Here, $\Delta s_{12}$, $\Delta s_{23}$, and $\Delta s_{31}$ are the edge lengths as depicted in Fig. 2(b). For the surfactant diffusion term, only the components in the direction of $\mathbf{p}$ are necessary. For later discussion of the interfacial source term, we also indicate the tangential, $\mathbf{t}$, component of the surface gradient in Fig. 2(b).

To obtain the surface surfactant gradient, we use the probing technique originally introduced by Udaykumar *et al*. [26]. Schematics for the general procedure for implementing the probing technique to compute the surface gradient of the surfactant in both the **p** and **t** directions are described in Fig. 3 (a) and (b), respectively. For example, to construct the surface gradient in the **p** direction at the center between nodes ① and ②, we construct a probe point $(x_{12}, y_{12}, z_{12})$ and define a probe distance *dl*, usually taken to be the grid size, in the normal direction $\mathbf{n}_{12}$; since we know that the interface is represented by the zero contour of the distance function field ($\phi$), we can locate the probe point to be on the interface where $\phi = 0$. We interpolate the surfactant concentrations $\Gamma_{in}$ and $\Gamma_{out}$ at the two points on either side of the interface from the probe point, i.e. $\mathbf{x}_{out} = (x_{out}, y_{out}, z_{out})$ and $\mathbf{x}_{in} = (x_{in}, y_{in}, z_{in})$ as shown in Fig. 3(a). Using these values the surface gradient in the **p** direction at point $\mathbf{x}_{12}$ can be computed by:

$$\nabla_{s\_p12} \Gamma = \frac{\Gamma_{out} - \Gamma_{in}}{2dl} \qquad (18)$$

The surface gradient at the other edges, i.e. $\nabla_{s\_p23}\Gamma$ and $\nabla_{s\_p23}\Gamma$ can be found in a similar way. The procedure to obtain the surface gradient in the **t** (tangential) direction is the same as above for the **p** direction except that we calculate values $\Gamma_{left}$ and $\Gamma_{right}$ on either side of the probe point along the element edge as shown in Fig 3(b). The surface gradient in the **t** direction at the point $\mathbf{x}_{12}$ is then:

$$\nabla_{s\_t12} \Gamma = \frac{\Gamma_{right} - \Gamma_{left}}{2dl} \qquad (19)$$

Thus for each Lagrangian element we store the values $\Gamma_{left}$, $\Gamma_{right}$, $\Gamma_{in}$, $\Gamma_{out}$, and $\Gamma_f$. We now describe how to transfer this Lagrangian information to the Eulerian grid since in parallel processing, communication of Eulerian quantities across subdomains is much simpler. We need to distribute the Lagrangian surface surfactant concentrations to the Eulerian grid by interpolation in the manner of [8]. The mass of surfactant can be found by integrating $\Gamma$ over the surface:

$$M_s = \int_{A_e} \Gamma \delta(\mathbf{x}-\mathbf{x}_f) dA \tag{20}$$

Here, δ(**x**-**x**$_f$) is a 3D Dirac distribution that is non-zero only when **x** = **x**$_f$. The discrete field can be computed by distributing the surface surfactant $\Gamma_f$ to the Eulerian grid as:

$$M_{sijk} = \sum_f \Gamma_f D_{ijk}(\mathbf{x}_f) \Delta A \tag{21}$$

where Δ$A$ is the element area and D$_{ijk}$ is the discrete Dirac distribution. For a given interface element located at **x**$_f$ = ($x_f$, $y_f$, $z_f$), we use the tensor product suggested by Peskin and McQueen [27]

$$D_{ijk}(\mathbf{x}_f) = \frac{\delta(x_f/h_x - i)\delta(y_f/h_y - j)\delta(z_f/h_z - k)}{h_x h_y h_z} \tag{22}$$

where $h_x$, $h_y$ and $h_z$ are the dimensions of an Eulerian grid cell and

$$\delta(r) = \begin{cases} \delta_1(r), & |r| \leq 1 \\ 1/2 - \delta_1(2-|r|), & 1 < |r| < 2 \\ 0, & |r| \geq 2 \end{cases} \tag{23}$$

and

$$\delta_1(r) = \frac{3 - 2|r| + \sqrt{1 + 4|r| - 4r^2}}{8} \tag{24}$$

Using Eqs. (21-24), the surface surfactant concentration at the center of each interface element can be distributed over a narrow width of 3 or 4 grid cells around the interface as per the usual Peskin immersed boundary technique. The total mass of the surfactant can be obtained by integrating the field $M_s$ over the entire domain. We divide this $M_s$ field by:

$$Q = \int_{A_e} \delta(\mathbf{x} - \mathbf{x}_f) dA \tag{25}$$

where the distribution field ($Q_{ijk}$) can be constructed in the same manner as that for $M_s$ using Eq. (21). The final surface surfactant field distribution $\Gamma(\mathbf{x})$ can then be obtained:

$$\Gamma(\mathbf{x}) = \frac{\int_{A_e} \Gamma \delta(\mathbf{x} - \mathbf{x}_f) dA}{\int_{A_e} \delta(\mathbf{x} - \mathbf{x}_f) dA} = \frac{M_{sijk}}{Q_{ijk}} \tag{26}$$

In order to check the accuracy of this procedure for distributing the surface surfactant concentration, we perform a simple benchmark test using a given surfactant concentration. For simplicity and clarity, we use 2D geometry. An initially circular interface with randomly distributed interface elements is placed at the center of the simulation domain (Fig. 4(a)). A sinusoidally varying surface surfactant concentration is assigned to the Lagrangian elements (i.e. at the center of each interfacial element). The Lagrangian values on the interface were then distributed using Eq. (21) and divided by $Q_{ijk}$ as in Eq. (25). Fig. 4(b) shows the computed surfactant concentration field resulting from Eq. (26). The interpolated surface surfactant concentration at the original Lagrangian interface node locations was plotted in Fig. 4(c). As can be seen from the figure, the interpolated surfactant values match very well with the given Lagrangian values at that location. At every time-step when the surface surfactant concentrations on the Lagrangian elements are updated, we also perform the distribution of $\Gamma$ to obtain the field distribution of the surfactant using Eq. (26).

Using Eq. (5), (11), and (12) the discretized form of the surfactant concentration conservation equation is:

$$\frac{(\Gamma A_e)^{n+1} - (\Gamma A_e)^n}{\Delta t} = \delta \Gamma_{D_e}^n + \dot{S}_\Gamma^n A_e^{n+1} \tag{27}$$

which can be rearranged to:

$$\frac{\Gamma^{n+1} - \Gamma^n \frac{A_e^n}{A_e^{n+1}}}{\Delta t} = \frac{\delta \Gamma_{D_e}^n}{A_e^{n+1}} + \dot{S}_\Gamma^n \tag{28}$$

If we define the area ratio between successive time steps as $A_r = A_e^n / A_e^{n+1}$, then the final form of the interfacial surfactant governing equation becomes:

$$\Gamma^{n+1} = \Gamma^n A_r + \Delta t \left[ \frac{\delta \Gamma_{D_e}^n}{A_e^{n+1}} + \dot{S}_\Gamma^n \right] \tag{29}$$

To summarize, the overall procedure for updating the surface surfactant concentration equation is as follows:

(i) First, the area ratio ($A_r$) is computed using the area of each interface element before ($A_e^n$) and after ($A_e^{n+1}$) interface advection (Eq. (10)) as ($A_r = A_e^n / A_e^{n+1}$).

(ii) The surface surfactant concentrations at the centers of the interfacial elements are distributed to the Eulerian grid using Eq. (26). At the same time, geometrical information such as normal, binormal, and tangential vectors as well as area and edge lengths are identified for each element.

(iii) The surface gradients in the **p** directions at the three edges of each element are computed using the probing technique from Eq. (18). Then the diffusion term in the surfactant conservation equation is evaluated using Eq. (17).

(iv) The source term in the right-hand-side of the interfacial surfactant conservation equation is computed using Eq. (6). Then, finally, the updated interfacial surfactant concentration at each interface element is found using Eq. (29)

2.4 Implementation of interfacial force with surfactant concentration

Following the original Front-tracking formulation of the surface tension force [6], the source term **F** in the momentum equation (Eq. (2)) can be given by

$$\mathbf{F} = \oint_C \sigma \mathbf{p}' dl \qquad (30)$$

Here $\sigma$ is the varying surface tension coefficient and is a function of the interfacial surfactant concentration. With surfactant on the interface, the surface tension coefficient can decrease. The surface tension coefficient can be modeled quite generally with $\sigma = \sigma(\Gamma)$ and several potential equations of state could be chosen [28]. A popular equation of state on thermodynamic grounds derived from Langmuir adsorption [17] for the case of dilute concentration, such that $\Gamma/\Gamma_\infty \ll 1$ is :

$$\sigma = \sigma_s + RT\Gamma_\infty \ln\left(1 - \frac{\Gamma}{\Gamma_\infty}\right) = \sigma_s \left[1 + \beta_s \ln\left(1 - \frac{\Gamma}{\Gamma_\infty}\right)\right] \qquad (31)$$

Here, $\sigma_s$ is the surface tension coefficient of clean surface, $R$ is the ideal gas constant, and $\beta_s = RT\Gamma_\infty/\sigma_s$ is defined as the elasticity number. In practice this model can be simplified to a linear model as:

$$\sigma = \sigma_s \left[1 - \beta_s \frac{\Gamma}{\Gamma_\infty}\right] \qquad (32)$$

and we use this in some numerical verifications and comparisons later. However, we note that nonlinear equations of state such as empirical models, [29] can be easily incorporated.

The vector $\mathbf{p}'$ in Eq. (30) is the binormal component of the product of the normal vector $\mathbf{n}_f$ and tangential vectors along the edges of the each element (Fig. 2(a)); *dl* is the edge length corresponding to $\Delta s_{12}$, $\Delta s_{23}$, and $\Delta s_{31}$ in Fig. 2(b). This formulation generates pulling forces along the interface edges thus generating a net inward force after integration around the entire surface. As indicated in Shin [24], this formulation can generate numerical parasitic currents since the discrete curl properties of the interfacial force do not match those of the discrete pressure gradient. A hybrid formulation of the curvature [8] has been proposed to overcome this difficulty and it was shown that parasitic currents can be suppressed significantly. However, this hybrid form of the surface tension force was developed for the case of a constant surface tension coefficient where only the normal component of the surface force is considered. Here, we revisit the hybrid idea and reformulate it for the case of varying surface tension coefficient. We divide the surface tension force into normal and tangential components as follows [12, 15]:

$$\mathbf{F} = \oint_C \sigma \mathbf{p}' dl = \int_{A_e} (\sigma \kappa \mathbf{n} + \nabla_s \sigma) dA = \int_{A_e} \sigma \kappa \mathbf{n} dA + \int_{A_e} \nabla_s \sigma dA = \mathbf{F}_n + \mathbf{F}_s \qquad (33)$$

Here, $\mathbf{F}_n$ represents the normal component, which can be described by the hybrid formulation for the discrete curvature $\kappa_H$ [8, 24].

$$\mathbf{F}_n = \int_{A_e} \sigma \kappa \mathbf{n} \delta(\mathbf{x} - \mathbf{x}_f) dA = \sigma \kappa_H \nabla I \qquad (34)$$

With varying surface tension coefficient, we computed $\sigma \kappa_H$ as a single field distribution:

$$\sigma \kappa_H = \frac{\mathbf{F}_L \cdot \mathbf{G}}{\mathbf{G} \cdot \mathbf{G}} \qquad (35)$$

where

$$\mathbf{F}_L = \int_{A_e} \sigma \kappa_f \mathbf{n}_f \delta(\mathbf{x}-\mathbf{x}_f) dA \qquad (36)$$

$$\mathbf{G} = \int_{A_e} \mathbf{n}_f \delta(\mathbf{x}-\mathbf{x}_f) dA \qquad (37)$$

Here $\mathbf{x}_f$ is a parameterization of the interface and $\kappa_f$ is twice the mean interface curvature obtained from the Lagrangian interface structure. The geometric information, unit normal, $\mathbf{n}_f$, and area of the interface element, $dA$, in $\mathbf{G}$ are computed directly from the Lagrangian interface and then distributed onto an Eulerian grid using the discrete Dirac distribution (similar to Eq. (21-24)).

To evaluate the tangential component of $\mathbf{F}$, $\mathbf{F}_s$, the surface gradient of the surface tension coefficient has to be identified:

$$\mathbf{F}_s = \int_{A_e} \nabla_s \sigma \delta(\mathbf{x}-\mathbf{x}_f) dA \qquad (38)$$

The surface tension gradient is further decomposed into its $\mathbf{p}$ and $\mathbf{t}$ components as in the evaluation of the surfactant diffusion term (see Eq. (16)) as follows:

$$\nabla_s \sigma = (\nabla_s \sigma)_p \mathbf{p} + (\nabla_s \sigma)_t \mathbf{t} \qquad (39)$$

A detailed description of the surface gradient vectors has been described in Fig. 2(b). The evaluation procedure for the example of a component at the center of the triangle edge $\overline{①②}$ was shown in Eqs. (18) and (19) using the probing method. The surface gradient for the other edges $\overline{②③}$ and $\overline{③①}$ can be obtained in a similar fashion. The distribution of the surface tension gradient to the Eulerian grid is a straightforward process similar to Eqs. (36) and (37). Each edge component of the surface tension gradient is distributed at the location of the edge center weighted by one third of the element area (Fig. 2(b)):

$$\begin{aligned}\mathbf{F}_s = & \left[ (\nabla_s\sigma)_{p12}\mathbf{p}_{12} + (\nabla_s\sigma)_{t12}\mathbf{t}_{12} \right]\delta(\mathbf{x}-\mathbf{x}_{12})A_{12} \\ & + \left[ (\nabla_s\sigma)_{p23}\mathbf{p}_{23} + (\nabla_s\sigma)_{t23}\mathbf{t}_{23} \right]\delta(\mathbf{x}-\mathbf{x}_{23})A_{23} \\ & + \left[ (\nabla_s\sigma)_{p31}\mathbf{p}_{31} + (\nabla_s\sigma)_{t31}\mathbf{t}_{31} \right]\delta(\mathbf{x}-\mathbf{x}_{31})A_{31} \end{aligned} \quad (40)$$

Using the hybrid formulation of the interface force minimizes parasitic currents also in the case of varying surface tension coefficient along the interface.

## 2.5 Sharp boundary condition for bulk surfactant concentration equation

The governing equation for bulk concentration is a straightforward convection diffusion equation except for the source term at the interface. In a classic Front-tracking approach, the interfacial source term would be embedded in the governing equation using a Dirac distribution at the interface. This method however requires an iterative approach to satisfy conditions at the phase interface and in addition can exhibit over or undershoots near the interface. Recently, ghost type methods [30-32] have become popular since boundary conditions can be directly applied to the phase interface. Since the ghost method enforces specific boundary conditions at the ghost node directly by extrapolation of values inside the computation domain, it does not require any additional iteration. Recently, Shin and Choi [25] developed a sharp energy method which prevents unstable or singular behavior of gradients when the interface comes too close to a grid node. Here, we modify this sharp energy method to account for the surfactant boundary condition for bulk concentration. The detailed procedure and implementation of the sharp boundary condition can be found in [25] and here we will briefly describe the key features for implementing the boundary condition for surfactant gradient Eq. (9).

Ghost nodes are only defined in Phase 1 since we allow bulk diffusion concentration only in phase 2 (Fig. 5). The concentration gradient at the phase interface can be approximated using a $1^{st}$ order gradient as follows:

$$\left.\frac{\partial C}{\partial n}\right|_f = \frac{C_{ghost} - C_{Xf}}{ds_{min}} = \frac{-\dot{S}_\Gamma}{D_{c2}} \qquad (41)$$

Here, $C_{ghost}$ is the ghost node value in Phase 1 and $C_{xf}$ is the surfactant concentration on the interface element; $ds_{min}$ is the minimum distance to the interface from a given ghost node which is obtained during the generation of the distance function field. The distance function field was directly computed by finding the minimum distance to the Lagrangian interface. (The detailed procedure can be found in [25].) The surfactant source term can be evaluated by Eq. (6) using the updated surface concentration field. Thus the concentration at the ghost node, i.e. adjacent to the phase boundary (see Fig. 5), can be computed as:

$$C_{ghost} = C_{Xf} - \frac{\dot{S}_\Gamma}{D_{c2}} ds_{min} \qquad (42)$$

A higher order version of this approach is readily possible and extension to bulk diffusion in both phases is straightforward.

# 3. RESULTS AND DISCUSSIONS

The hydrodynamic aspects of our method and its treatment of interface dynamics have been thoroughly tested in previous simulation studies of, for example, drop oscillation, bubble rise, drop impact, Rayleigh Taylor instability, Faraday instability etc. [7-8, 22-24]. In this section, we focus on validation of the accuracy of our newly formulated algorithm for surfactant transport. We present several benchmark tests which demonstrate individually the accuracy of surfactant mass conservation (Section 3.1), surfactant diffusion on the interface (Section 3.2), the sharp boundary condition for bulk surfactant (Section 3.3), and the surface tension force formulation (Section 3.4). The overall performance of our algorithm is tested in Section 3.5 for the case of drop deformation in simple shear flow with surfactant, and in Section 3.6 for the influence of surfactant on capillary bridge dynamics and rupture.

## 3.1 Simulation of a uniformly expanding sphere

We first test the case of a uniformly expanding sphere in order to check the accuracy of the calculation of surfactant mass conservation via Eq. (26). We are especially interested in the performance after the interface has undergone multiple interface reconstruction steps. We place a sphere with radius 0.5 in a 2.5×2.5×2.5 box with a uniform initial surfactant concentration on the interface of unity ($\Gamma_{f\_ini} = 1$). There is no surfactant absorption or desorption in this test and the sphere expands at a constant velocity of 0.01 in the radial direction. Thus the surfactant concentration on the interface will depend solely on the surface area ratio during expansion. Since the surfactant mass on the interface should be preserved during expansion, the exact dimensionless surface surfactant concentration can be computed by:

$$\Gamma^*(t) = \frac{\Gamma(t)}{\Gamma_\infty} = \frac{A_{ini}}{A(t)} \frac{\Gamma_{ini}}{\Gamma_\infty} = \frac{A_o}{A(t)} \Gamma^*_{ini} \qquad (43)$$

where $A_{ini}$ and $\varGamma_{ini}$ are the initial surface area and initial surface surfactant concentration, respectively. We plot the variation of the surface surfactant concentration during expansion in Fig. 6 along with the analytical result from Eq. (43). For this simulation, the domain was discretized by a $64^3$ grid and the initial sphere diameter was discretized by only 12 grid cells. Thus this represents a rather rigorous low resolution test. As can be seen in Fig (6), the agreement is excellent even at this low resolution despite a total of 1000 interface reconstruction steps during the simulation. The interface was reconstructed every 20 time-steps in order to assess that accuracy is maintained during frequent reconstruction.

## 3.2 Stationary droplet with diffusion

One of the crucial aspects of our formulation lies in the method for surfactant diffusion on the interface. Here we simulate surfactant diffusion on a spherical surface with radius 0.5 placed in a 2.5×2.5×2.5 box with a non-uniform surfactant distribution $\varGamma = (1-\cos\theta)/2$ where $\varGamma$ is non-dimensionalized by $\varGamma_\infty$. and $\theta$ is the polar angle measured from the axis of symmetry. The exact solution for this problem is:

$$\varGamma(\theta,t) = \frac{1}{2}\left(1 + e^{-\frac{2t}{D_s R^2}} \cos\theta\right) \qquad (44)$$

Here, $D_s$ is the diffusion coefficient which was set to unity. We show the variation of the surface surfactant concentration over time compared with this exact solution in Fig. 7(a). Again we discretize the domain using a rather low resolution $64^3$ mesh and as can be seen in Fig. 7(a) the surface surfactant diffusion is computed accurately compared to the exact solution even at such low resolution.

We tested this same case at higher resolution to obtain the convergence behavior with increasing mesh resolution. In Fig. 7(b), the $L_1$ norm defined by Eq. (45) was plotted for three different resolutions of $64^3$, $128^3$, and $256^3$:

$$L_1 = \frac{1}{N}\left(\sum_{i=1}^{N}|\Gamma_{cal} - \Gamma_{exact}|\right) \qquad (45)$$

Here, *i* represents the discretization index, *N* the total number of mesh points; $\Gamma_{cal}$ is our simulation result and $\Gamma_{exact}$ is the exact value from Eq. (44). As can be seen in Fig. 7(b), we obtain between 1st and 2nd order convergence with increasing grid resolution.

### 3.3 Bulk concentration diffusion and mass transfer test

Another key feature in our method is the sharp boundary condition for bulk surfactant transport. As discussed in Section 2.4, the source term for the surface surfactant equation is enforced via a Neumann condition at the phase boundary. Thus the correct implementation and accuracy of the boundary condition Eq. (9) is important for the proper evaluation of the bulk transport given by Eq. (7).

To validate the accuracy of the sharp boundary condition for the bulk concentration, we conduct a test case similar to that of Muradoglu and Tryggvason [17]. An initially clean drop with radius 0.5 is placed in 5×5×5 box. The initial bulk concentration $C_\infty$ was fixed to be constant. We only allow adsorption onto the surface from the bulk by including the source term $\dot{S}_\Gamma = k_a C_s$ thus ensuring that surfactant will transfer from the bulk to the interface. The approximate analytical solution to this problem provided by Muradoglu and Tryggvason [17] is:

$$\frac{C - C_\infty}{C_\infty} = \frac{k_a\sqrt{\pi D_s t}/D_s}{1 + \frac{\sqrt{\pi D_s t}}{R}\left(1 + \frac{k_a R}{D_s}\right)} \frac{r}{R} erfc\left(\frac{r - R}{2\sqrt{D_s t}}\right) \qquad (46)$$

The cross sectional bulk concentration is shown in Fig. 8 compared with the approximate analytical solution Eq. (46) demonstrating that the current formulation treats the source term accurately.

## 3.4 Marangoni migration of a droplet

Next we perform a test of the accuracy of the surface tension formulation for the case of drop migration due to variation in the surface tension coefficient. As in thermocapillary migration, an initially stationary droplet moves due to a surface tension gradient, here due to a surfactant concentration gradient which is prescribed as a function of the axial location. Marangoni stress caused by a non-uniform surface tension force along the interface will drive the stationary droplet in the direction of lower surfactant concentration [33]. We place a spherical droplet with radius 0.5 at the bottom of the simulation domain (center 0.6R from the bottom wall). The domain size is 5R×5R×10R in each direction and periodic boundary conditions were applied in the horizontal directions, i.e. in the *x* and *y* directions, with no-slip walls at both top and bottom. A constant density and viscosity of 0.2 and 0.1 were used for both phases, respectively. The surfactant concentration on the interface was specified as a function of height as:

$$\frac{\Gamma}{\Gamma_\infty} = \frac{z}{L_z} \tag{47}$$

Here, $L_z$ is the box size in the *z*-direction. The surface tension coefficient is described using the linear model (Eq. 32):

$$\sigma = \sigma_s\left(1 - \beta_s \frac{\Gamma}{\Gamma_\infty}\right) = \sigma_s\left(1 - \beta_s \frac{z}{L_z}\right) \tag{48}$$

This case corresponds exactly to thermocapillary migration of a viscous droplet in a linear ambient temperature gradient with equal conductivity in both phases and steady terminal velocity of the rising droplet and can be estimated using the YGB linear model [34] for small Reynolds and Marangoni numbers. The rise velocity for the YGB model is:

$$V_{YGB} = \frac{2\sigma_s \beta_s R}{L_z(9\mu_G + 6\mu_L)} \qquad (49)$$

Fig. 9 shows the non-dimensional rise velocity of the droplet based on the YGB model. The rise velocity converges with increasing grid resolution and matches relatively well with the theoretical rise velocity from the YGB model. The discrepancy in rise velocity is argued by Muradoglu and Tryggvason [17] to be due to the effect of the finite box size in the horizontal direction. Alhendal *et al*. [33] demonstrated that the rise velocity is also affected by Reynolds number (Re = $V_{rise}R/\nu$). In our simulation the rise velocity slightly underestimates the YGB model even though horizontally periodic boundary conditions were used to minimize side wall effects. However our solution is comparable to the results of [17].

## 3.5 Drop deformation under shear

In this section, more realistic conditions are tested in order to ascertain the overall accuracy of the surfactant model. We simulate drop deformation in a constant shear flow produced at the top and bottom boundaries of the domain as shown in Fig. 10. A spherical droplet with initial radius $4.88 \times 10^{-4}$ m is placed at the center of the simulation domain. A box size of 8R×8R×4R is used in the *x*, *y*, and *z* directions respectively considering the recommendation of Komrakova *et al*. [35] in order to minimize boundary effects. Periodic conditions are applied in the horizontal directions (*x* and *y*) and a simple shear flow is prescribed at the top and bottom of the domain with velocity $\dot{\gamma}z$ where $\dot{\gamma}$ represents the characteristic shear rate. Simulations were performed until steady state was reached. The density for both phases is a constant 1030 kg/m³ and the surface tension coefficient for a clean surface is $\sigma_s = 3.12 \times 10^{-2}$ kg/s² with the elasticity number $\beta_s$ as 0.8. The viscosity for Phase 1, inside the droplet, is set to 0.4 kg/m/s and that for Phase 2 is specified by the viscosity ratio λ with $\mu_2 = \lambda\mu_1$ where λ=3.335. This case corresponds to the simulation in Fig. 4 of Feigl *et al*. [14] and Fig. 9 of de Jesus *et al*. [19].

First, we compare our simulation results with experiment and the numerical result of Feigl *et al*. [14]. We simulated drop stretching for the insoluble case with a shear strain $\dot{\gamma}$ of 15.97 which corresponds to a capillary number of Ca ($= R\mu_1\dot{\gamma}/\sigma_s$) of 0.1. We simulated both the case with and without surfactant. The final steady shape of the deformed droplet can be seen in Fig. 10 along with the surfactant concentration distribution. Fig. 10 (a) shows the final droplet shape with the effect of surfactant. The droplet becomes elongated in the shear direction to form an ellipsoidal shape. We also plot the major axis of the ellipsoid vs. the rotation angle ($\theta_{\text{rot}}$). In order to quantitatively compare with experiments and results from other numerical solutions, the width (*W*), depth (*B*), and length (*L*) of the ellipsoid is defined in Fig. 10(a).

As can be seen in Fig. 10, the surfactant concentration is higher at the tip of the elongated ellipsoid and lower in the less elongated width and depth directions. The rotation angle with surfactant is much less than with a clean surface as seen in Fig. 10(b). The rotation angles for the present simulations were 0.26 and 0.65 rad with and without surfactant respectively. The corresponding rotation angles from the experiments were 0.27 and 0.61 rad. Before comparing the detailed evolution of the length, width, and depth of the droplet, we tested grid convergence with increasing grid resolution. Fig. 11 shows the length of the droplet vs. time. The time has been non-dimensionalized by shear strain such that $t^* = \dot{\gamma}t$. As can be seen in Fig. 11, the simulation results converge with increasing resolution. We also plot the surfactant mass with time in Fig. 11(b). The results demonstrate that the surfactant mass conservation is very accurate for the converged solution. At a resolution of $128^3$, the simulation can be considered to be a converged solution.

The variation in length (*L*), depth (*B*), and Width (*W*) for our current method was compared with experimental results (dashed line) and results from other numerical simulations (dash dot line) in Fig. 12. Our simulations show very similar results to that of Feigl *et al*. [14] who couple a boundary integral method for the interface velocity, phase-tracking for the interface evolution, and the finite element method for surfactant concentration. Experimental values are very similar to both simulations except for the length of the droplet with surfactant. A more complete study for various *Ca* and viscosity ratios is necessary but here our main intention was to check the validity of the surfactant formulation. Overall, our current surfactant formulation compares well with experiment and other simulations.

We further investigate the droplet shear problem by allowing adsorption and desorption at the phase interface with soluble surfactant. We use a desorption coefficient of 1.6 which corresponds to a Biot number of $Bi$ $(= k_a / \dot{\gamma})$ of 0.1 with no adsorption in order to check the desorption effect independently. Then we set the adsorption coefficient to 1.6 with zero desorption to check the adsorption effect alone. In both cases, the diffusion coefficient for the bulk phase ($D_{c2}$) is set to 0.01 m$^2$/s. The evolution of the length, depth, and width of the droplet with changing desorption and adsorption coefficient is shown in Fig. 13. As can be seen in this figure, the drop dimensions tend to follow the insoluble surfactant case (red lines) at first but gradually converge to the final shape of the clean surface as expected. Fig. 14 shows the surfactant concentration distribution at different times for the case of desorption coefficient equal to 1.6. For this case, the initial droplet was placed in a clean environment ($C_\infty=0$) with a maximum packing concentration on the interface ($\Gamma_\infty$) of 1.0. Other conditions are the same. The absolute magnitude of the surfactant concentration decreases with time. Thus the length of the droplet decreases with larger rotation angle. It is interesting to note that the surfactant concentration is still high at the tip of the ellipsoid even while the variation along the surface is significantly reduced.

On the other hand, the interface evolution shows a drastic change when we consider adsorption alone. In this case, the initial bulk concentration was set to 1.5 throughout the domain and the maximum packing concentration on the interface ($\Gamma_\infty$) was set to 2. As can be seen in Fig. 16, the interface is stretched continuously in the shear direction. Once again a high concentration of surfactant is located at the tip of the elongated surface. The overall magnitude of the surfactant concentration continually grows due to adsorption from the ambient bulk fluid. We also plot the bulk concentration field at the final times for both the desorption and adsorption test cases in Fig. 15 and 17, respectively. As can be seen from the figures, the bulk phase receives surfactant from the surface for non-zero desorption. For non-zero adsorption, the bulk concentration is depleted relative to the ambient value but there is nevertheless a slight increase of the concentration at the center of the rotation. This is due to accumulation of the surfactant at the minimum shear location.

We also perform a simulation of drop deformation at higher shear rate to check the algorithm's performance in handling topological changes. The simulation conditions are the same as in the previous desorption and adsorption tests except a higher *Ca* of 0.5 and lower elastic number ($\beta_s = 0.2$) is used for this simulation. Here we consider both desorption and adsorption at the interface. As can be seen in Fig. 18, the interface stretches to the shear direction with rotation. A high concentration of surfactant is again observed at the tips of the ellipsoidal surface and the average surface concentration increases over time. After stretching in the shear direction the interface becomes dumbbell shaped with narrow bridges forming between the side bulges. These bridges thin and finally rupture resulting in the formation of a satellite drop. It is interesting to note that the surfactant concentration is still high at the tips of the small droplets after breakup. This simulation demonstrates that our method is able to handle complex topological changes due to Marangoni stresses and surfactant transport.

3.6 Annular, surfactant-laden flows

A case of importance to chemical processes, pharmaceutical and petrochemical industries is that of two-phase, counter-current annular flow inside of a vertical tube. Here we simulate such a flow for three scenarios: (a) a clean air/water interface, (b) an interface with insoluble surfactant, and (c) an interface with soluble surfactant and diffusion of surfactant in the bulk water. The tube has a radius of 16.2 mm. The inner wall of the tube is initially coated uniformly by a thin liquid film that is 1 mm thick. Although the tube has a length of 132 mm, we impose periodic boundary conditions in the vertical direction in order to be able to simulate the flow behavior in much longer tubes. The three simulations were performed on 3×3×6 cores of a parallel computing cluster using domain decomposition and the MPI message passing interface. The cylindrical air/water interface within the solid tube and the domain decomposition are shown in Fig 19. Local calculations on a 32×32×64 mesh were performed in parallel on each of the 54 subdomains. Thus the global mesh resolution was 96×96×384.

Initially all fluid velocities are zero though both fluids are acted upon by gravity, and,

additionally, we impose an upward volumetric force of 28.5 N/m$^3$ to the air in the core of the tube in order to establish a counter-current flow situation. We use values of 1.205 kg/m$^3$ and 1000 kg/m$^3$ for the densities of the air and water, respectively, and 1.78×10$^{-5}$ kg/m·s and 0.902×10$^{-3}$ kg/m·s for the absolute viscosities of air and water, respectively. The surface tension coefficient of the clean interface is $\sigma_s = 7.28 \times 10^{-2}$ kg/s$^2$.

Snapshots at t = 1 s from calculations performed for three different cases are shown in Fig. 20: (a) Case 1, clean interface, (b) Case 2, insoluble surfactant, and (c) Case 3, soluble surfactant with bulk transport in the water film. For Case 2, the interface contains insoluble surfactant with an initially uniform concentration, $\Gamma_0 = 5 \times 10^{-6}$ mol/m$^2$. The surface diffusion coefficient is $D_s = 6.48 \times 10^{-4}$ m$^2$/s, the maximum packing interfacial concentration is $\Gamma_\infty = 10^{-5}$ mol/m$^2$ and we set the adsorption and desorption coefficients, $k_a$ and $k_d$ to zero. For the coupling of interface surfactant concentration to variation in interfacial surface tension we use the Langmuir relation, Eq. (31), with Elasticity number, $\beta_s$=0.35. In Case 3, we allow desorption of surfactant from the interface to the bulk via Eq. (6) with desorption coefficient, $k_d$=1.6 s$^{-1}$. The bulk surfactant transport in the water is then calculated by Eq. (7) with bulk diffusion coefficient, $D_c$=10$^{-3}$ m$^2$/s. All other parameters are the same as in Case 2.

Fig. 21(a) corresponds to the image in Fig. 20(b) and shows the distribution of surfactant on the interface for the insoluble case. Fig. 21(b) shows the interfacial surfactant distribution and Fig. 21(c) the bulk surfactant distribution on a mid-plane slice corresponding to the soluble case shown in Fig. 20(c). In comparing these three cases, the presence of surfactant impacts the dynamics of the air/water interface through a complex interplay of viscous shear stresses and Marangoni stresses due to local variation in the interfacial surfactant concentration. For Case 2, Marangoni stresses dominate due to the higher interface surfactant concentration gradients thus tending to smooth interfacial waves and inhibit droplet formation compared to Case 1 for the clean interface and Case 3 where interfacial surfactant concentrations are lower due to desorption from the interface into the bulk. A plot of the time evolution of interfacial surface area in Fig. 22 indicates that wave amplitudes and especially droplet formation are lower in the insoluble surfactant Case 2 compared to the clean and soluble cases corroborating what is observed in Fig. 20. The surface area of the soluble case 3 remains higher with sustained droplet generation and entrainment. Longer run times would, of course, be needed to fully characterize the spatio-temporal evolution of the film.

To go beyond this brief qualitative study of the effect of soluble and insoluble surfactant on interface dynamics, a thorough parametric study of these cases would be necessary to be able to provide quantitative results and identification of the precise roles of flow phenomena such as tip streaming, ligament and droplet formation as well as characterization of wave types. Such a dedicated study will be the subject of future work.

# 4. CONCLUSION

We propose a general procedure for soluble surfactant transport in the context of the Level Contour Reconstruction Method (LCRM). The LCRM is designed to benefit from advantages associated with the Front-tracking and Level Set methods to accurately track the phase evolution as in Front-tracking as well as efficiently handle topological changes as in the Level Set method. Additionally, the numerical schemes we describe for the surfactant transport processes retain the ability to be implemented on massively parallel distributed computing architectures for very high resolution simulations.

We describe an approach for calculating the surface gradient of the surfactant concentration in order to evaluate surfactant diffusion on the interface. A hybrid surface tension force formulation is extended to account for the varying surface tension coefficient which generates tangential surface forces. This is necessary in order to properly evaluate the gradient of the surface tension coefficient in the tangential direction. For soluble surfactant the interfacial source term is accounted for in the bulk phase concentration equation using a sharp boundary condition in order to correctly apply the Neumann boundary condition at the interface.

These newly devised formulations are individually tested for accuracy. Surfactant mass conservation on an expanding sphere is verified. Surfactant diffusion on the interface was compared with an exact solution and grid convergence verified. The sharp boundary method for the source term implementation in the bulk transport equation was compared with an approximate theoretical solution. The surface tension formulation for varying surface tension coefficient was tested for Marangoni drop translation. Finally, the overall performance of this surfactant model was compared with experiment and results from other simulations of drop deformation in simple shear flow. All cases considered here showed very good results comparable with either exact solutions, previous numerical work or experiments. Furthermore, we assessed the effect of desorption from and adsorption to the interface in drop deformation with shear flow. All benchmark tests compared well with existing data thus providing confidence that our adapted LCRM formulation for surfactant advection and diffusion is accurate and effective in three-dimensional multiphase flows.

As a demonstration for industrially relevant applications, we performed large scale parallel calculations of two-phase annular film flow in the counter-current flow regime both with and without the presence of surfactant and both with and without surfactant solubility. These simulations showed qualitative features expected in such a complex flow: wave and ligament formation, droplet detachment and entrainment and flooding. A full in depth study of annular film flow is, however, beyond the scope of this article and is the focus of upcoming work.

# ACKNOWLEDGEMENTS

This work is supported by (1) the Engineering & Physical Sciences Research Council, United Kingdom, through the MEMPHIS program grant (EP/K003976/1), (2) the Basic Science Research Program through the National Research Foundation of Korea (NRF) funded by the Ministry of Science, ICT and future planning (NRF-2014R1A2A1A11051346) and (3) by computing time at the Institut du Developpement et des Ressources en Informatique Scientifique (IDRIS) of the Centre National de la Recherche Scientifique (CNRS), coordinated by GENCI (Grand Equipement National de Calcul Intensif).

# FIGURE CAPTIONS

Fig. 1   General description of the LCRM.

Fig. 2   Description of geometrical information for an individual interface element (a) interface normal, binormal and tangential vectors at the edges as well as interface normal at the center of the element (b) direction of the surface gradient along the edge of the interface element.

Fig. 3   Data exchange between Lagrangian interface points and Eulerian grid (a) randomly distributed Lagrangian interface elements (b) distributed Eulerian surfactant distribution (c) comparison between given value at the Lagrangian interface and interpolated value with distributed surfactant field.

Fig. 4   Probing technique to evaluate surface gradient (a) **p** directional gradient (b) **t** directional gradient.

Fig. 5   Sharp boundary method for implementing source term in bulk concentration equation.

Fig. 6   Expansion of spherical surface for surfactant mass conservation test.

Fig. 7   Benchmark tests for surfactant diffusion on the interface (a) comparison of simulated results with exact solution at different times (b) grid convergence test.

Fig. 8   Benchmark test for bulk diffusion simulation. Numerical results were compared with an approximate theoretical solution.

Fig. 9   Marangoni drop migration test to check the accuracy of the interfacial surface tension force for varying surface tension coefficient.

Fig. 10  Drop deformation under simple shear stress (a) with surfactant case (b) clean case.

Fig. 11  Drop deformation test (a) grid convergence for the length of the major axis (b) surfactant mass conservation with different grid resolution.

Fig. 12  Comparison with experiment and results from another simulation (Feigl *et al*. [14]) (a) length (b) depth (c) width vs. non-dimensional simulation time.

Fig. 13  Effect of desorption and adsorption coefficient for the drop deformation in simple shear (a) length (b) depth (c) width vs. non-dimensional simulation time.

Fig. 14   Interface evolution with desorption from the interface. Surfactant concentration on the interface has been color contoured for different times with indicator for major axis direction

Fig. 15   Bulk concentration corresponds to the time (d) in Fig. 14.

Fig. 16   Interface evolution with adsorption from the interface. Surfactant concentration on the interface has been color contoured for different times with indicator for major axis direction

Fig. 17   Bulk concentration corresponds to the time (d) in Fig. 16.

Fig. 18   Interface evolution at higher *Ca* number of 0.5 with both adsorption and desorption from the interface. Surfactant concentration on the interface has been color contoured.

Fig. 19   Simulation domain with subdomain decomposition into 3x3x6=54 cores for parallel computation. Also shown are the initial cylindrical air/water interface (blue) within the solid tube (grey).

Fig. 20   Snapshots of the air/water interface at t = 1s for (a) case 1, clean interface, (b) case 2, interface with insoluble surfactant and (c) case 3, soluble surfactant with bulk surfactant diffusion. The outline of the solid tube (grey) is also shown.

Fig. 21   Snapshots of the air/water interface at t = 1s colored by (a) interface surfactant concentration for the insoluble case 2 corresponding to Fig. 20b, (b) interface surfactant concentration for the soluble case 3 corresponding to Fig. 20c and (c) bulk surfactant concentration on a slice through a mid-plane for the soluble case 3 corresponding to Fig. 20c.

Fig. 22   Time evolution of the interfacial surface area for the three cases shown in Fig. 20.

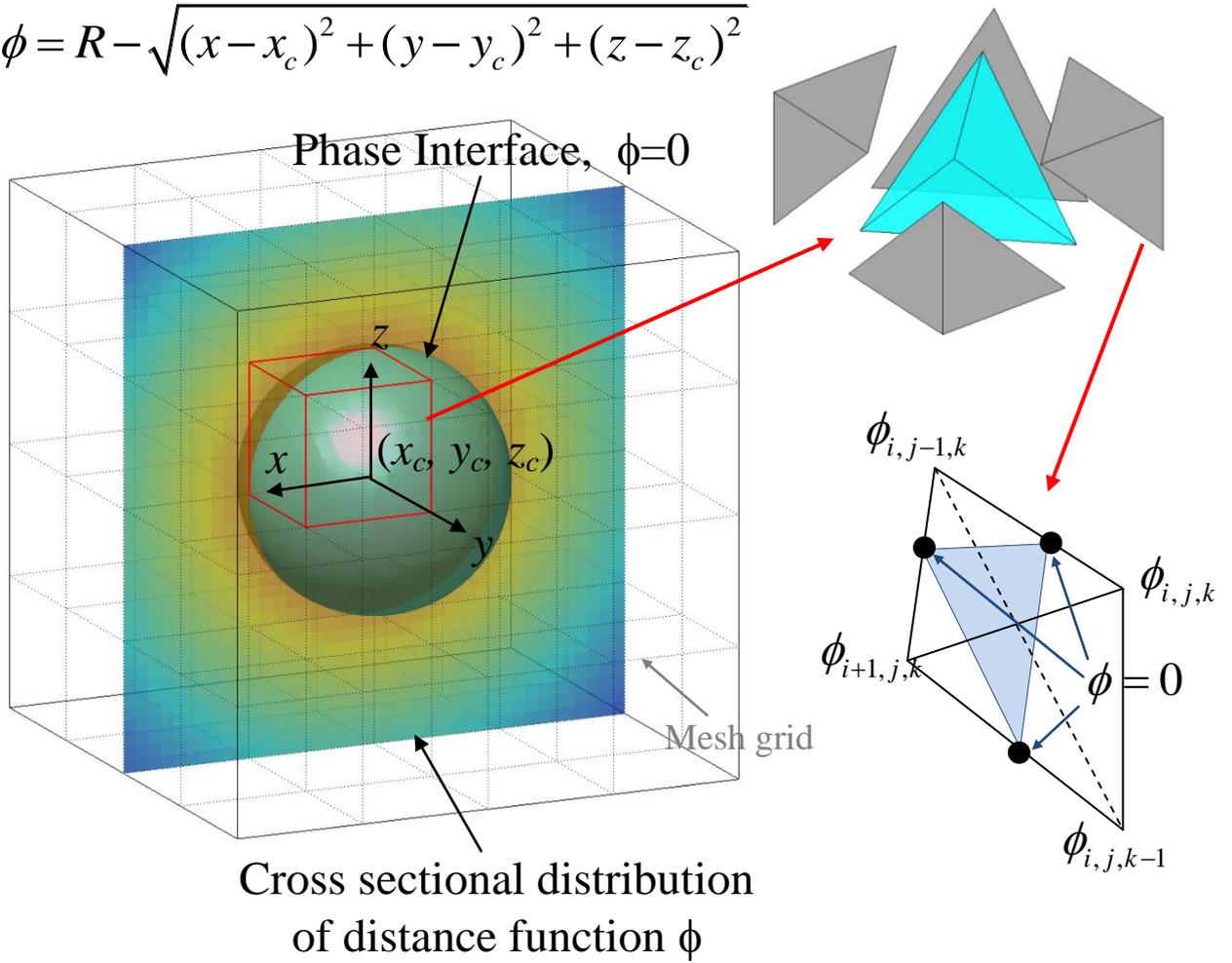

**Fig. 1**

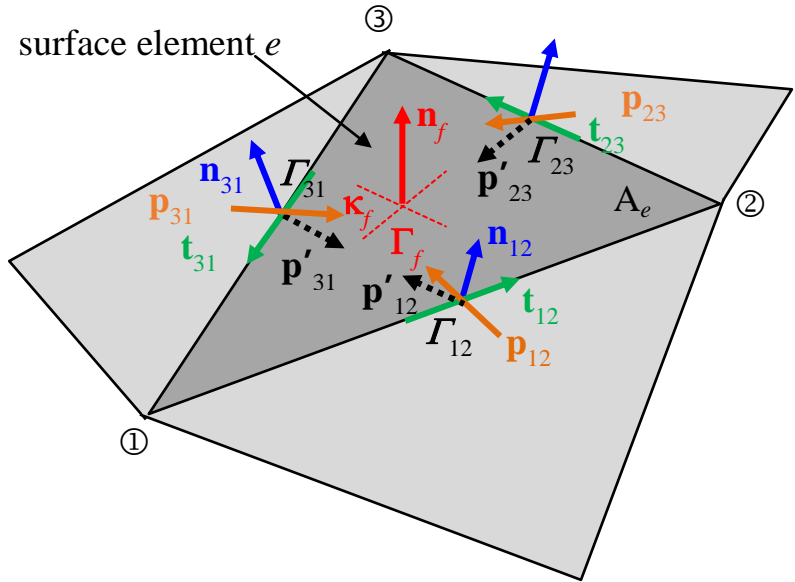

(a)

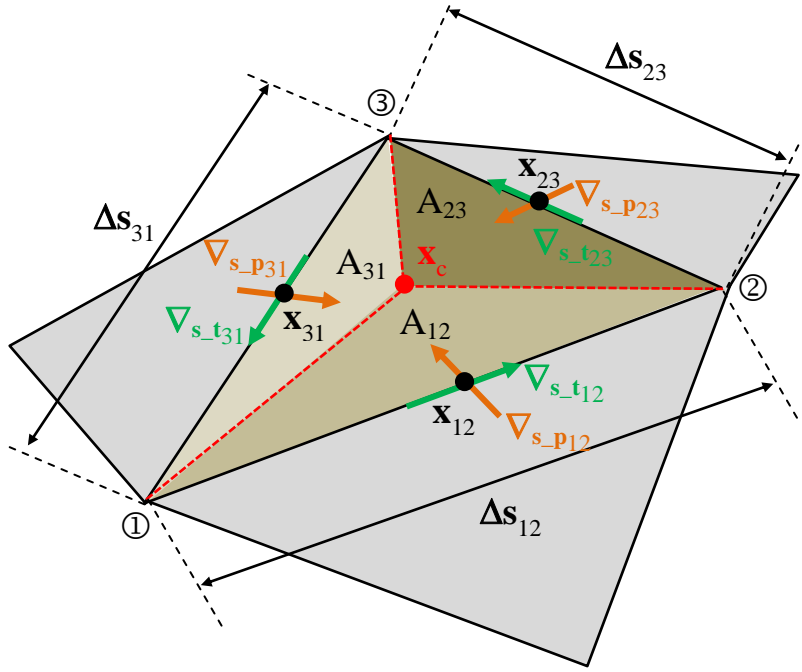

(b)

**Fig. 2**

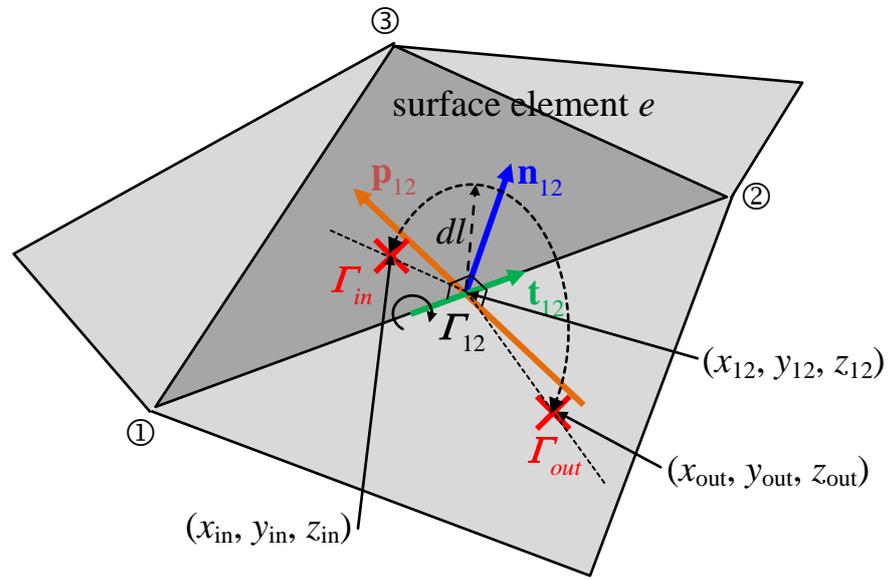

(a)

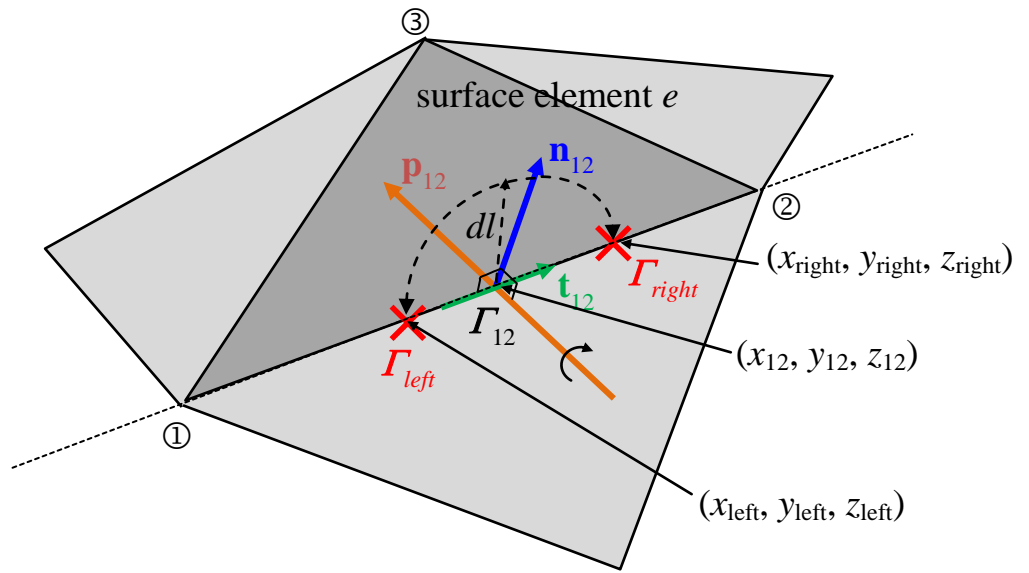

(b)

**Fig. 3**

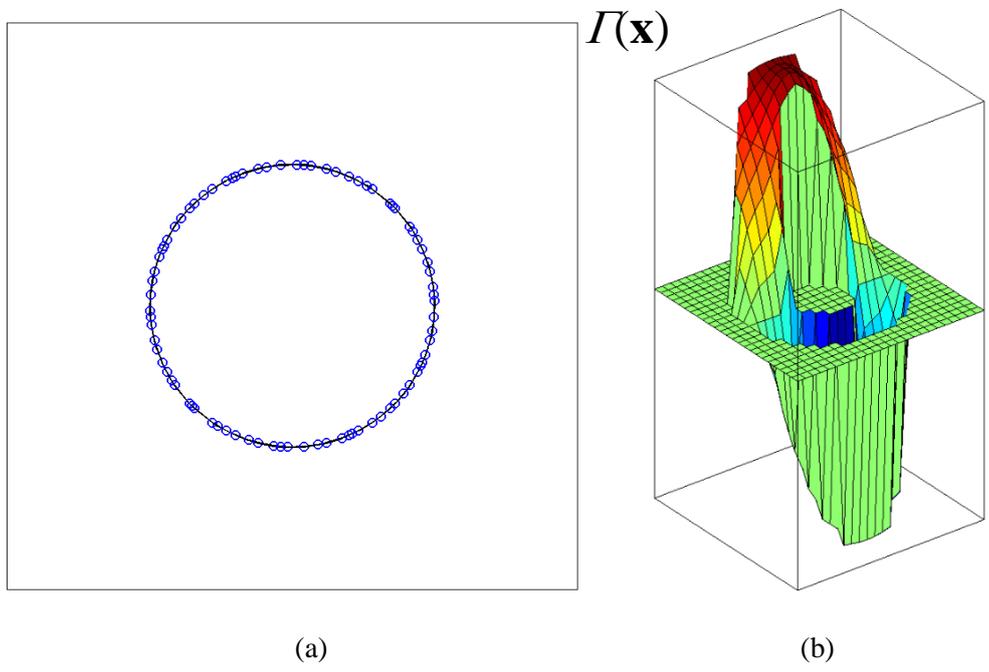

(a)          (b)

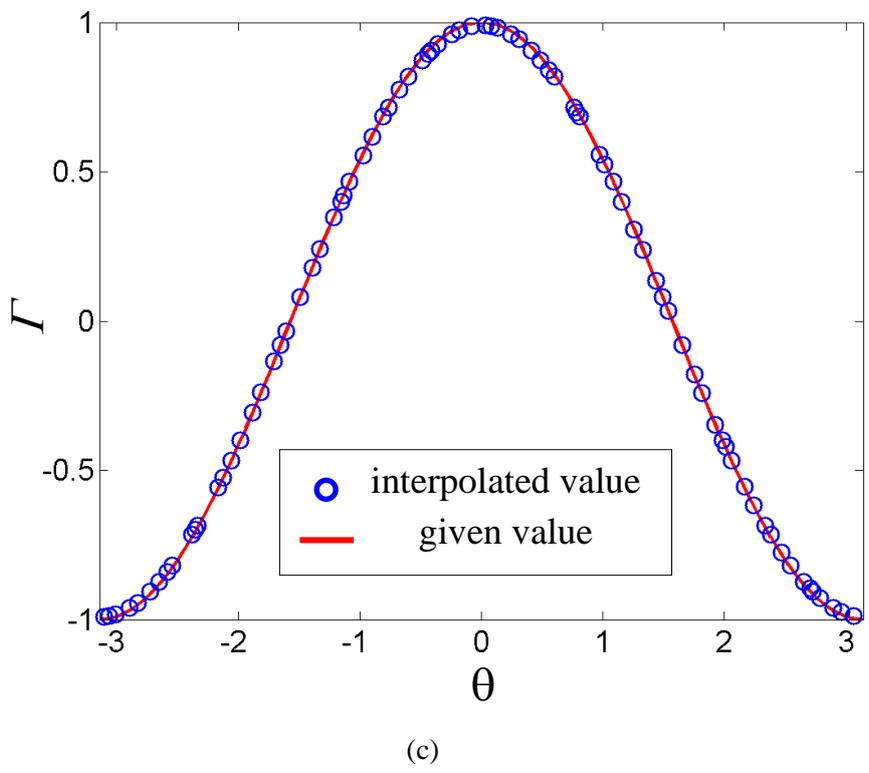

(c)

**Fig. 4**

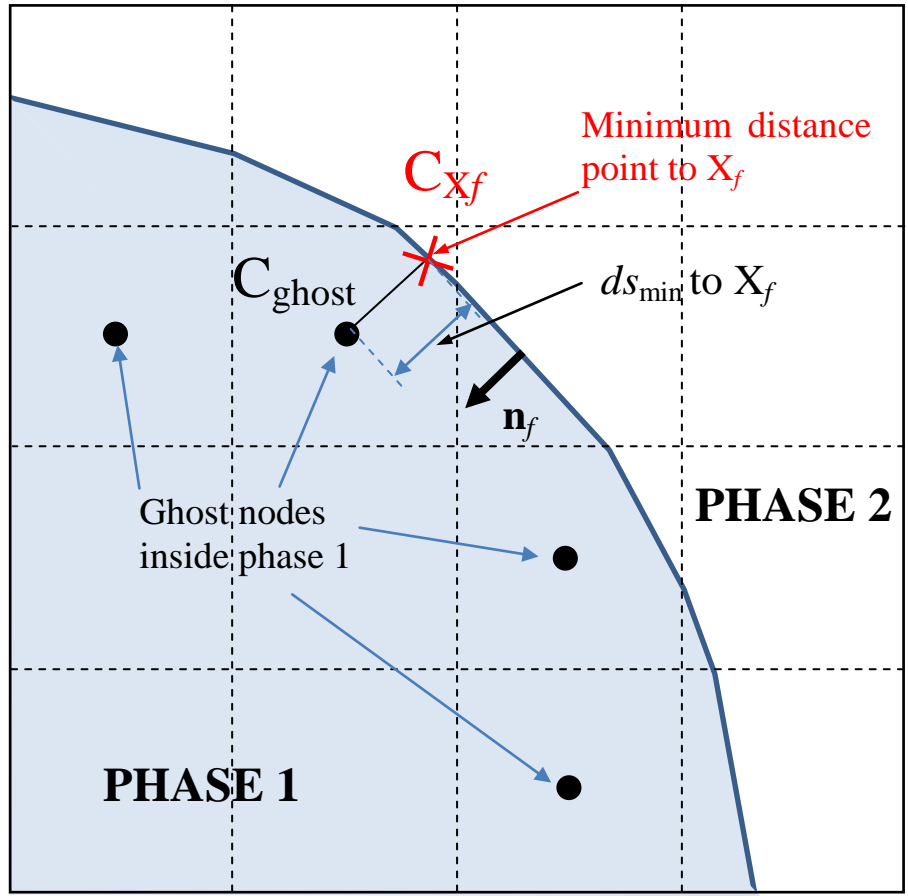

**Fig. 5**

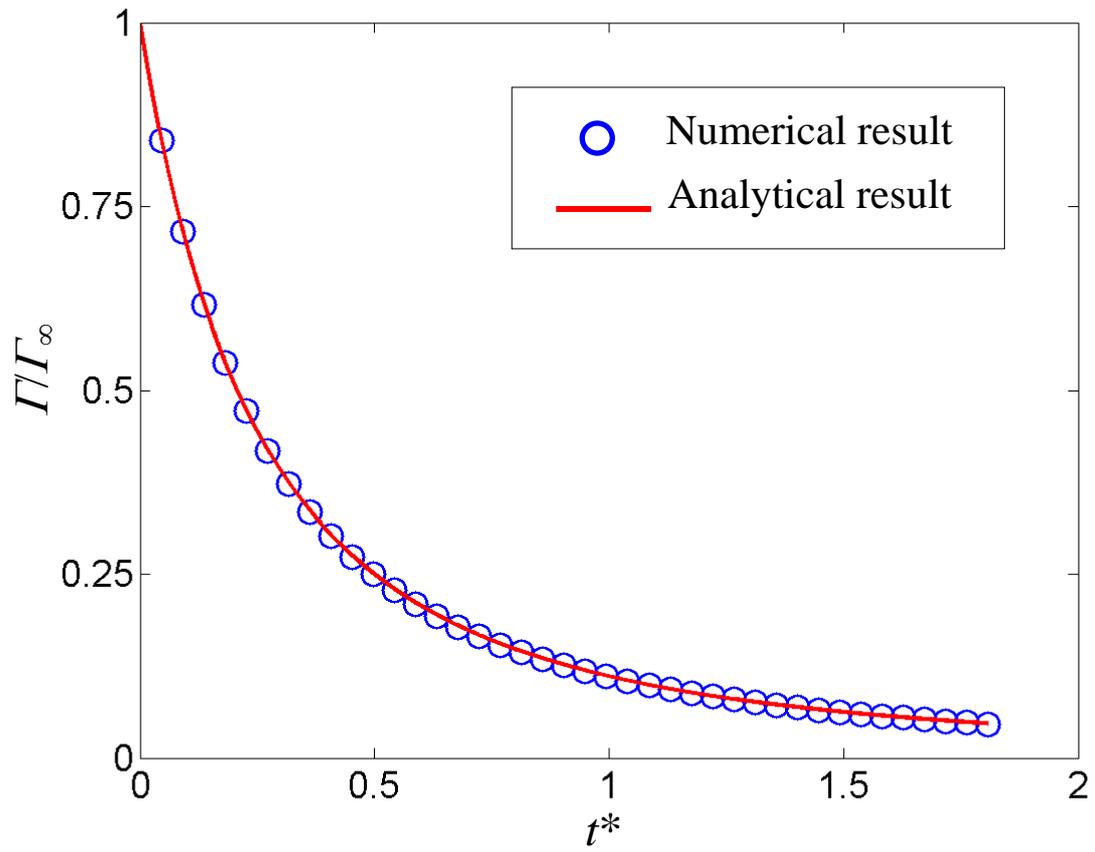

**Fig. 6**

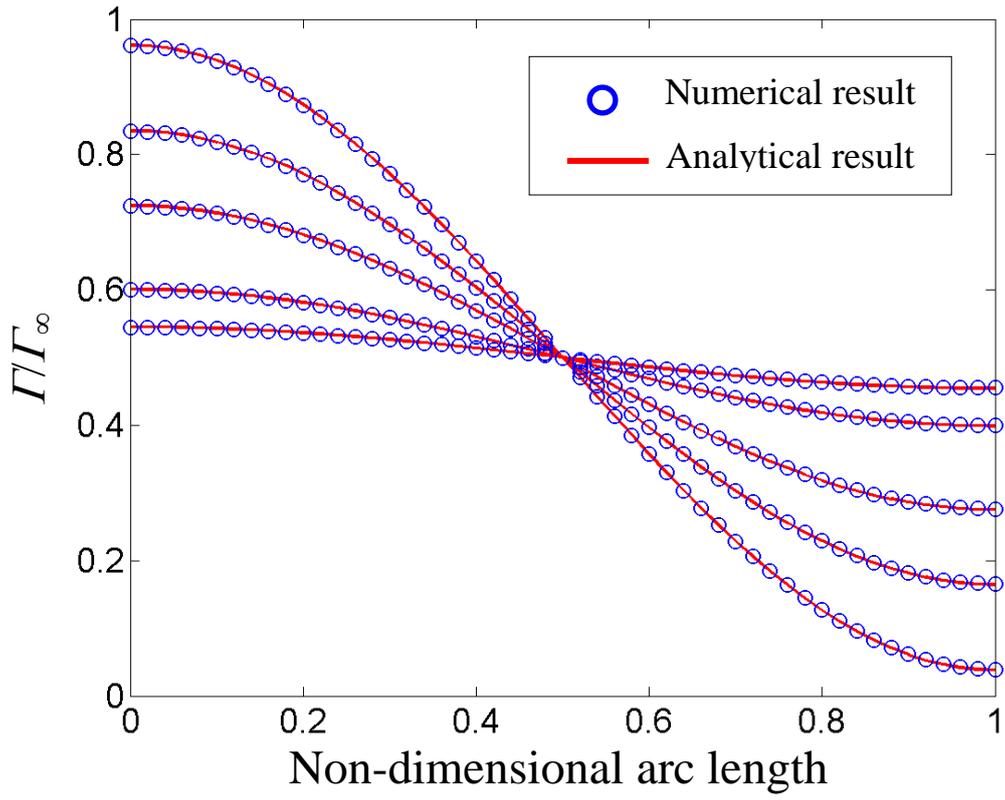

(a)

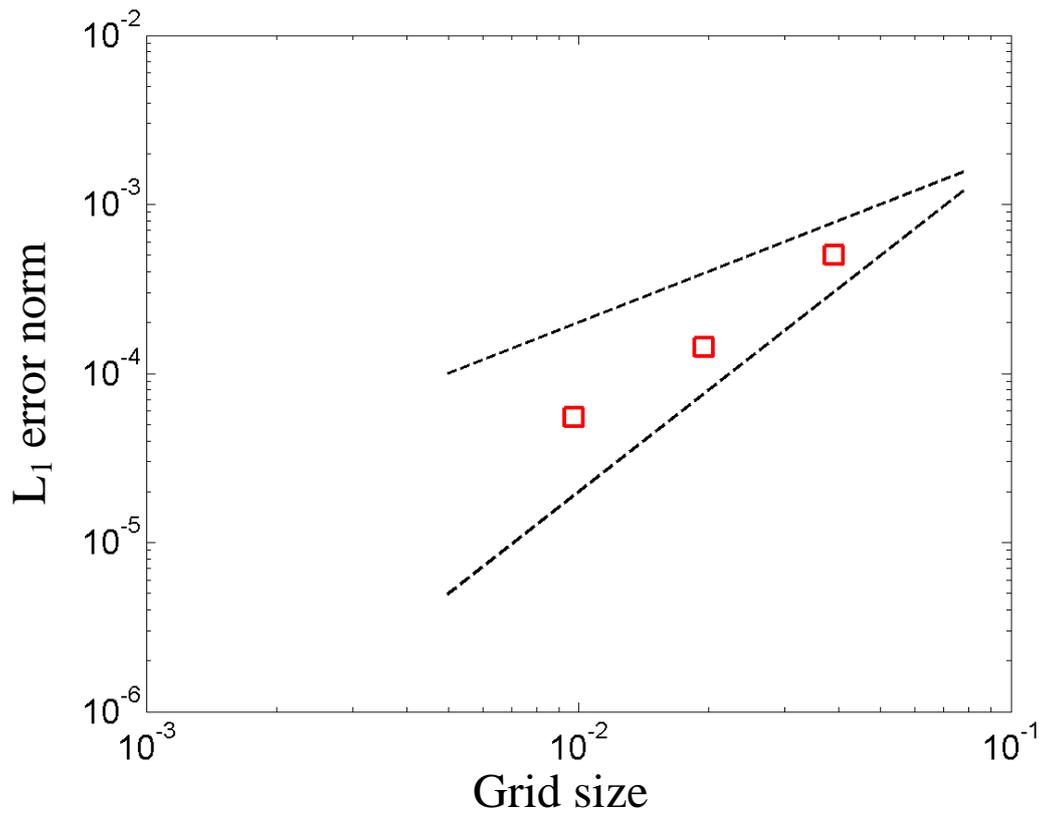

(b)

**Fig. 7**

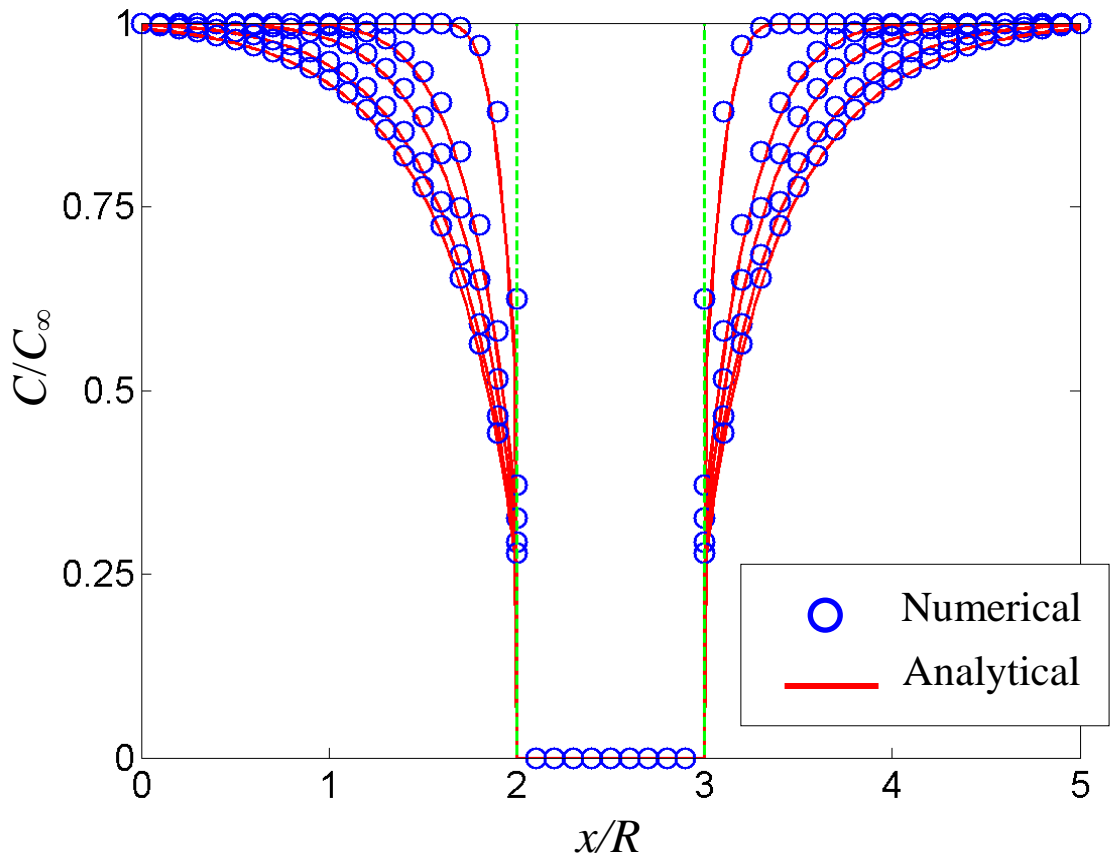

**Fig. 8**

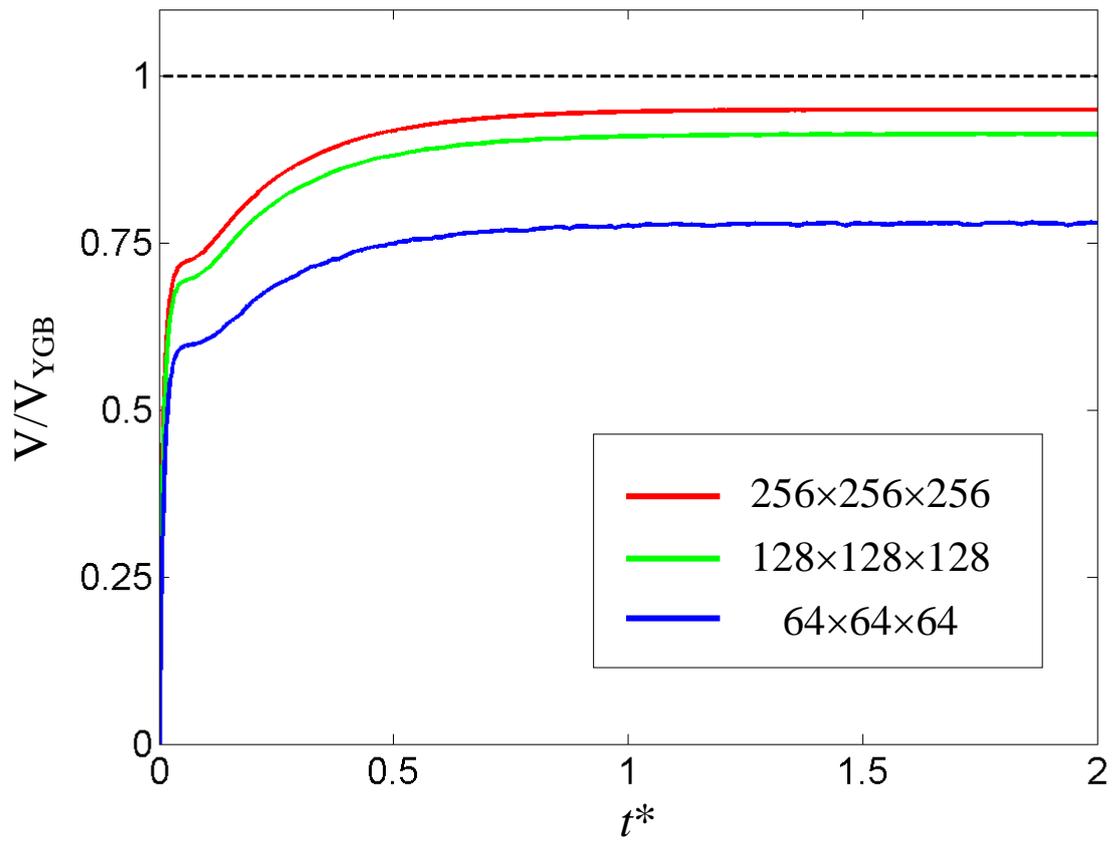

**Fig. 9**

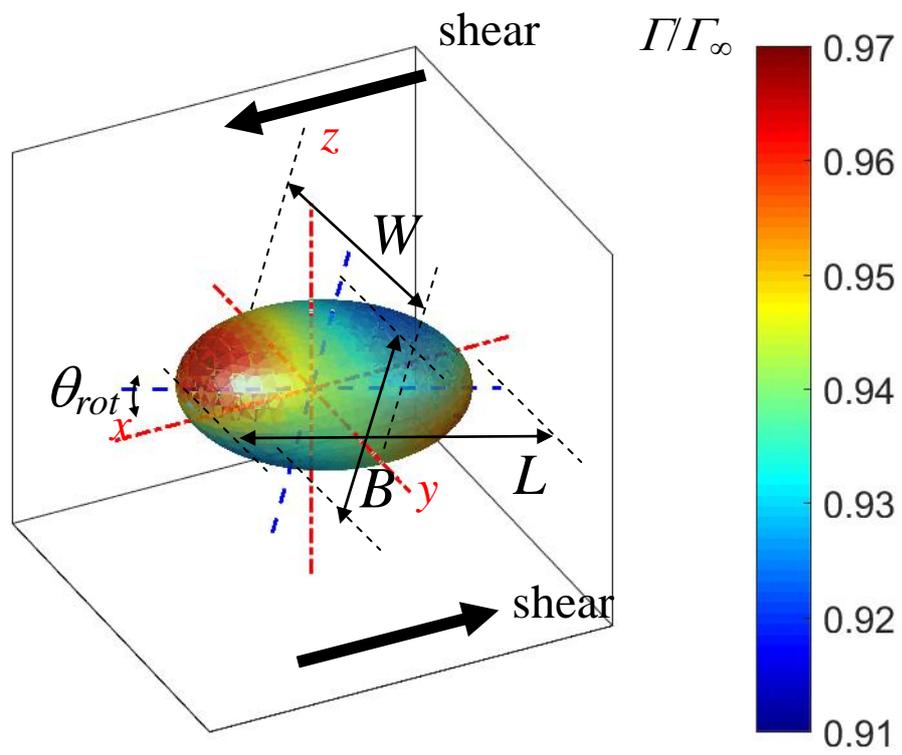

(a)

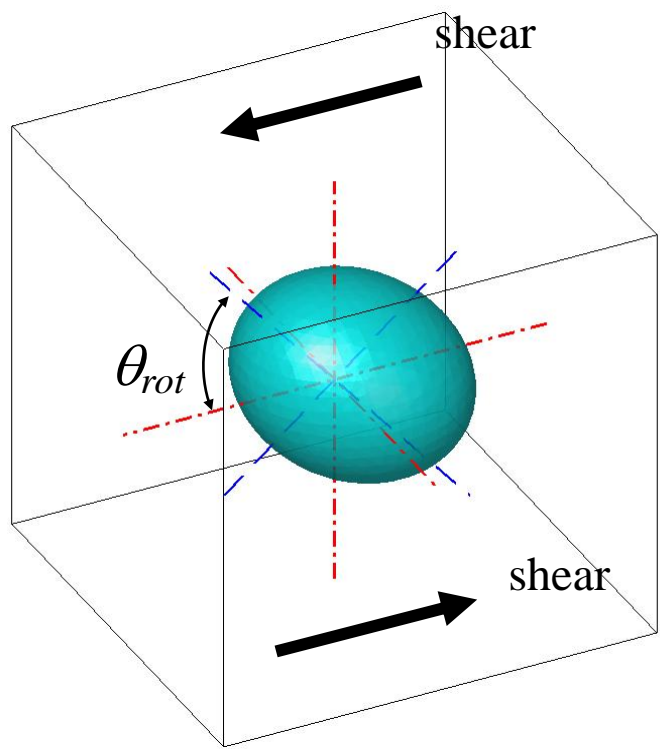

(b)

**Fig. 10**

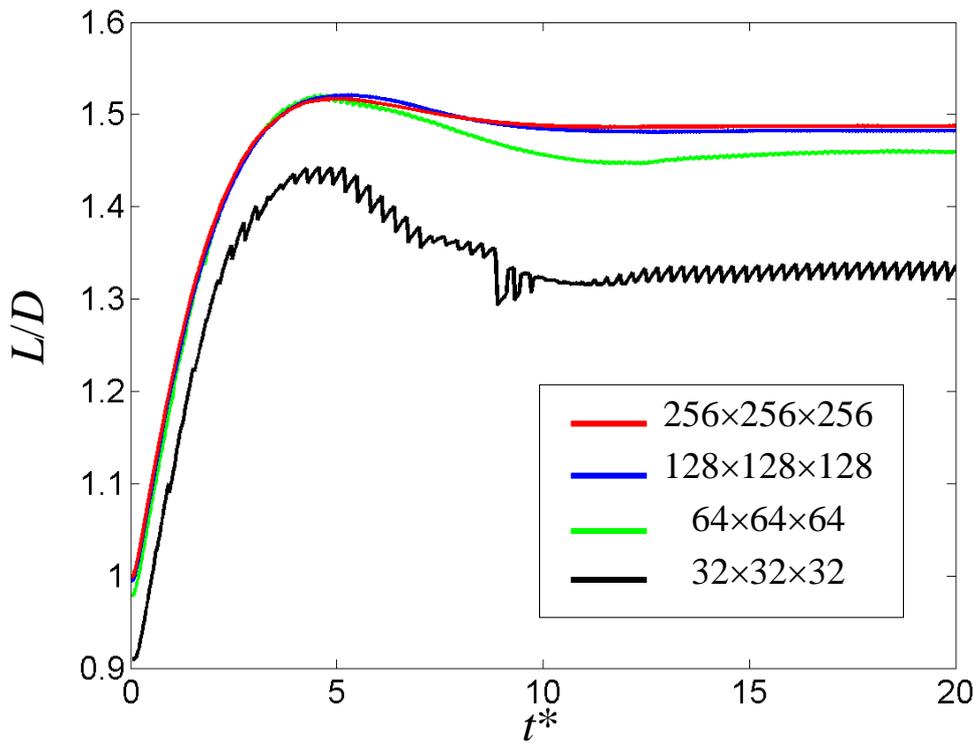

(a)

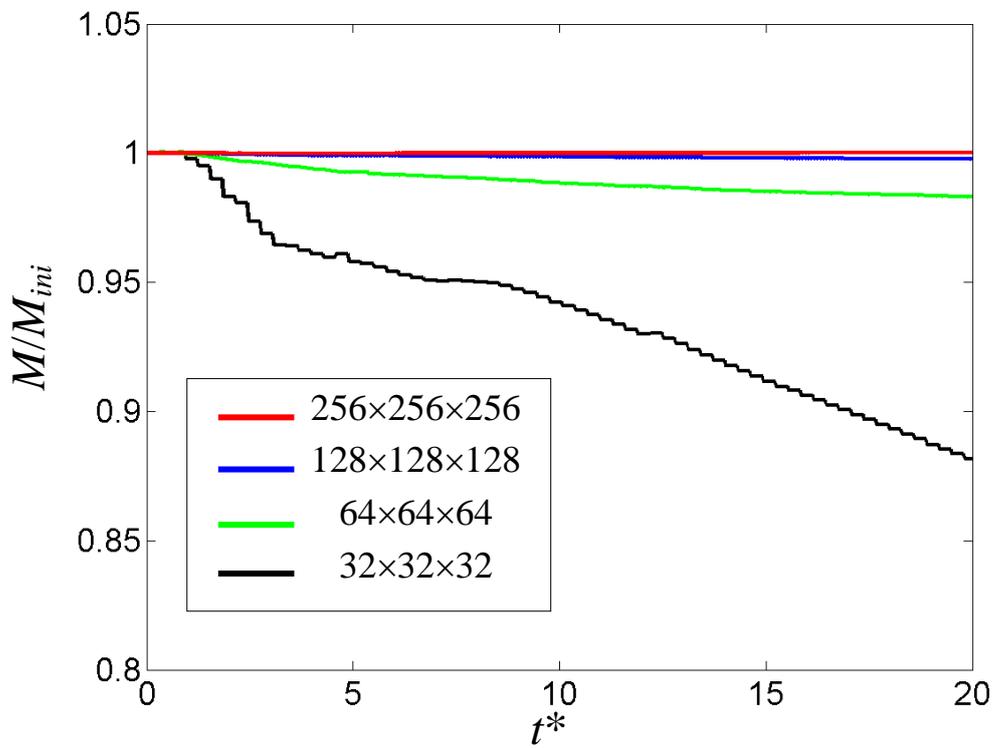

(b)

**Fig. 11**

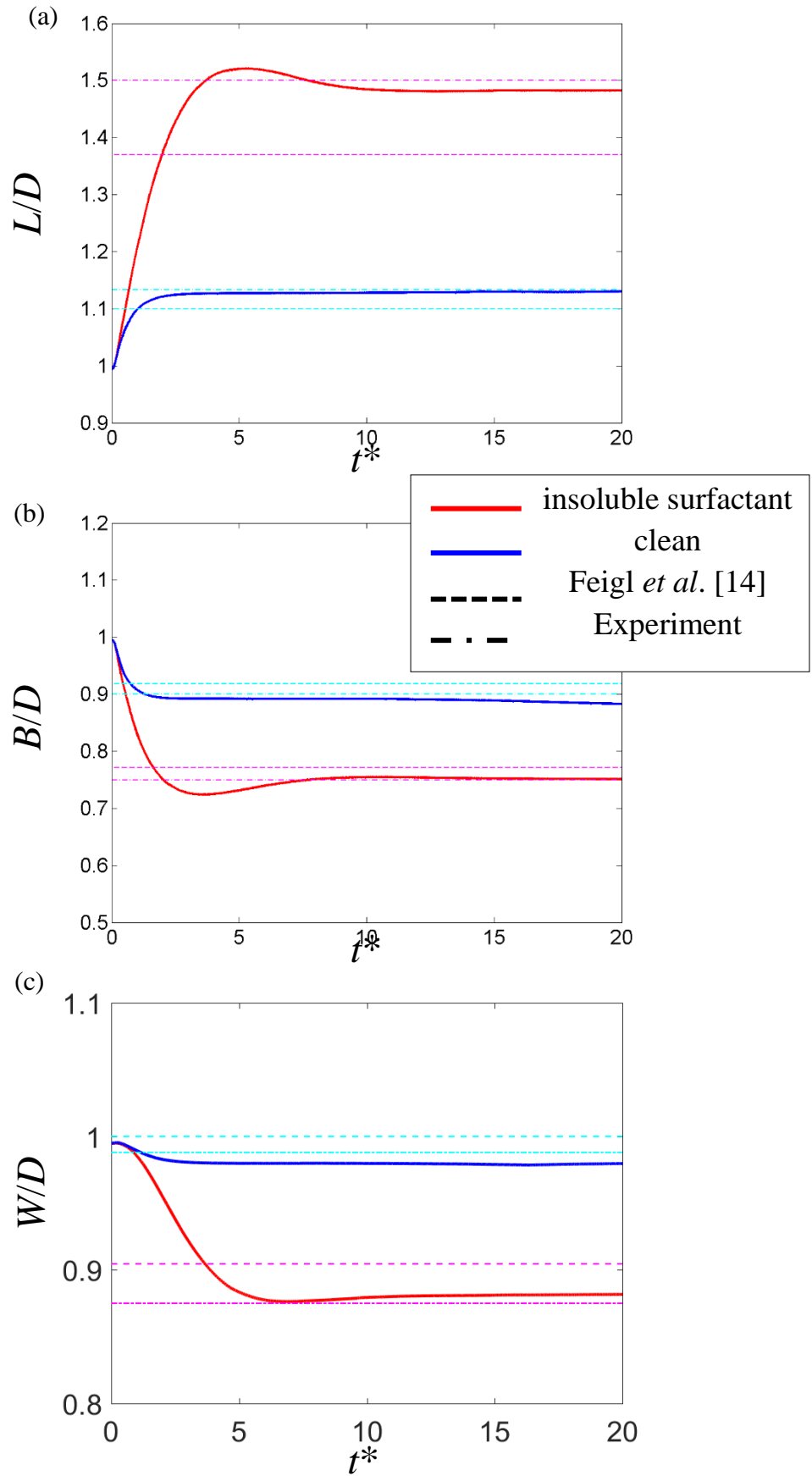

**Fig. 12**

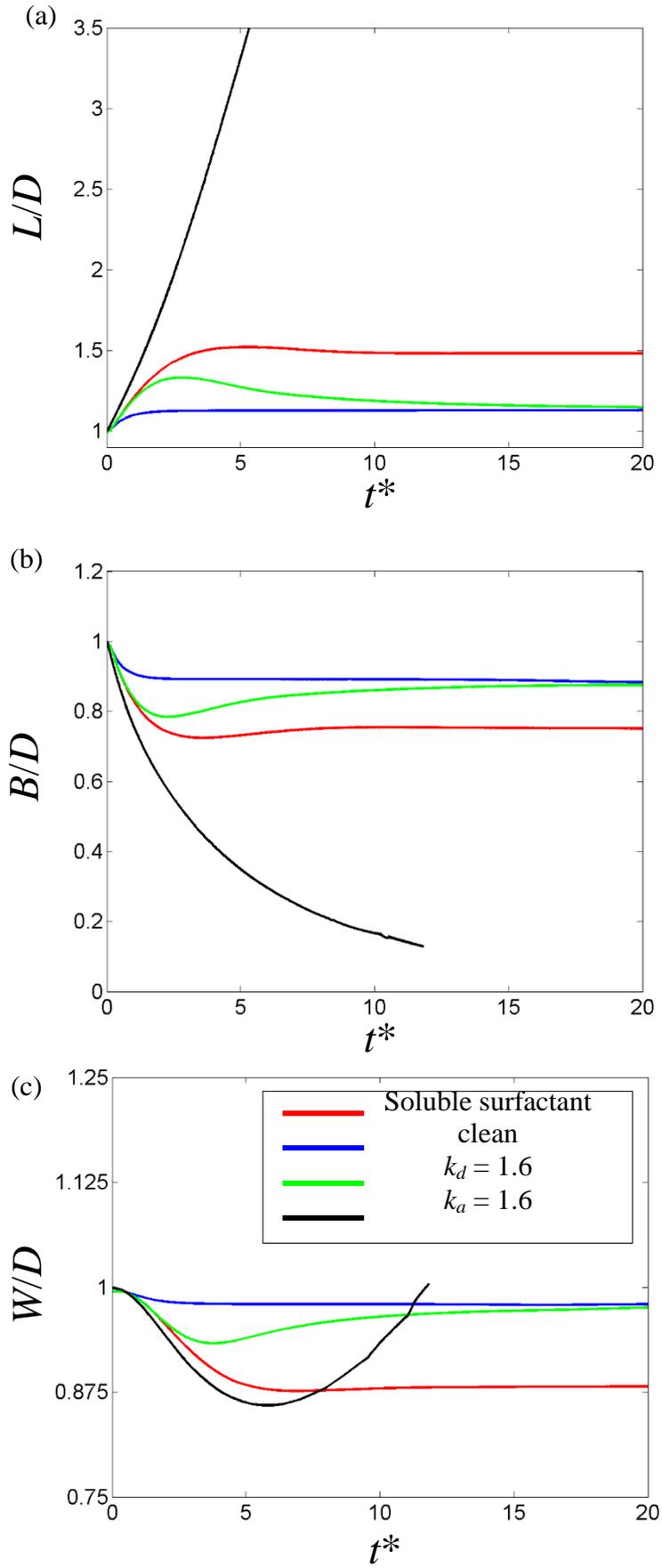

**Fig. 13**

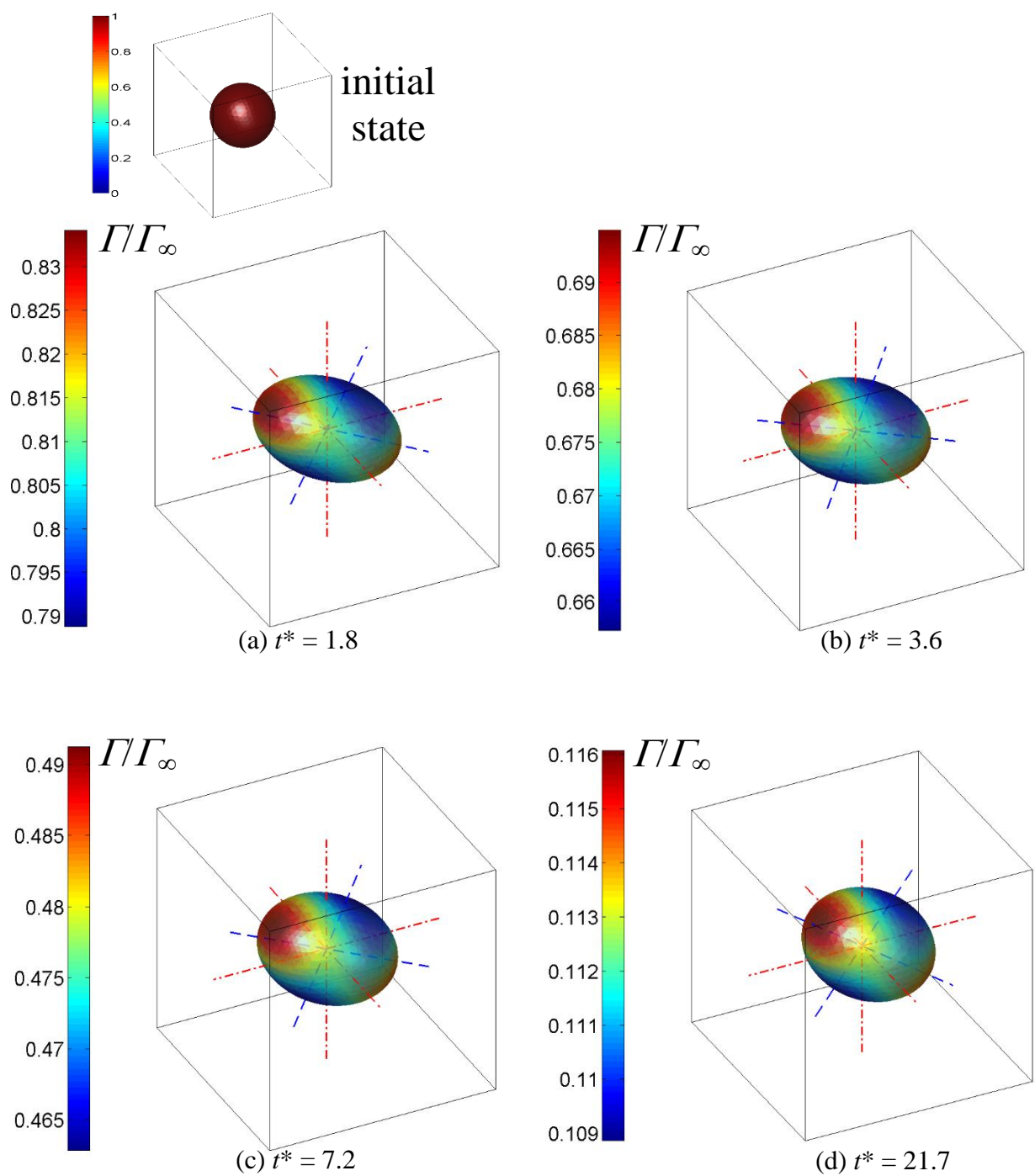

**Fig. 14**

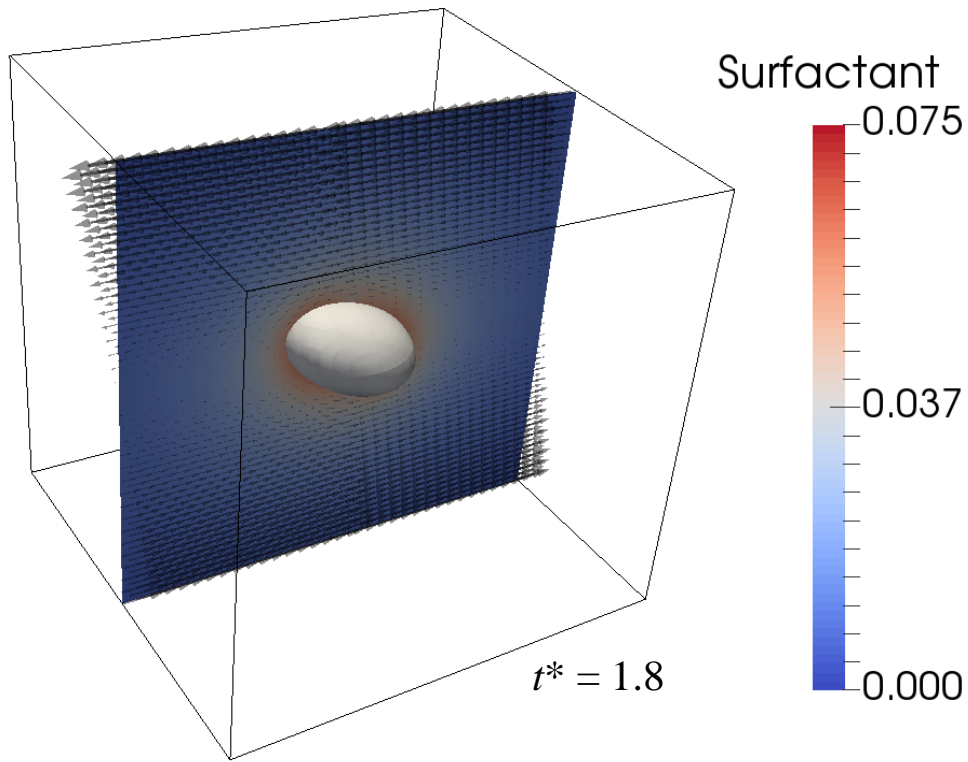

**Fig. 15**

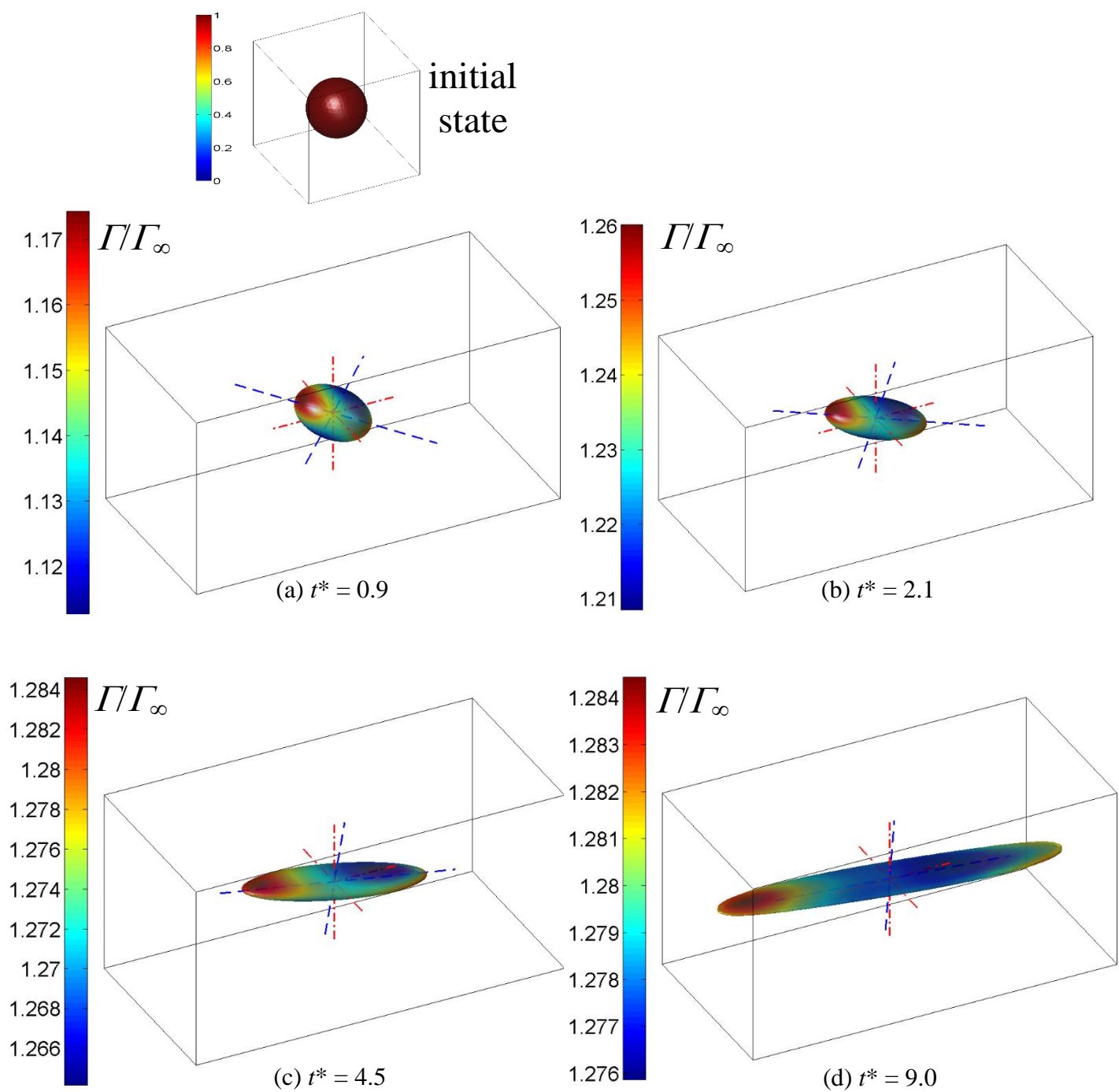

**Fig. 16**

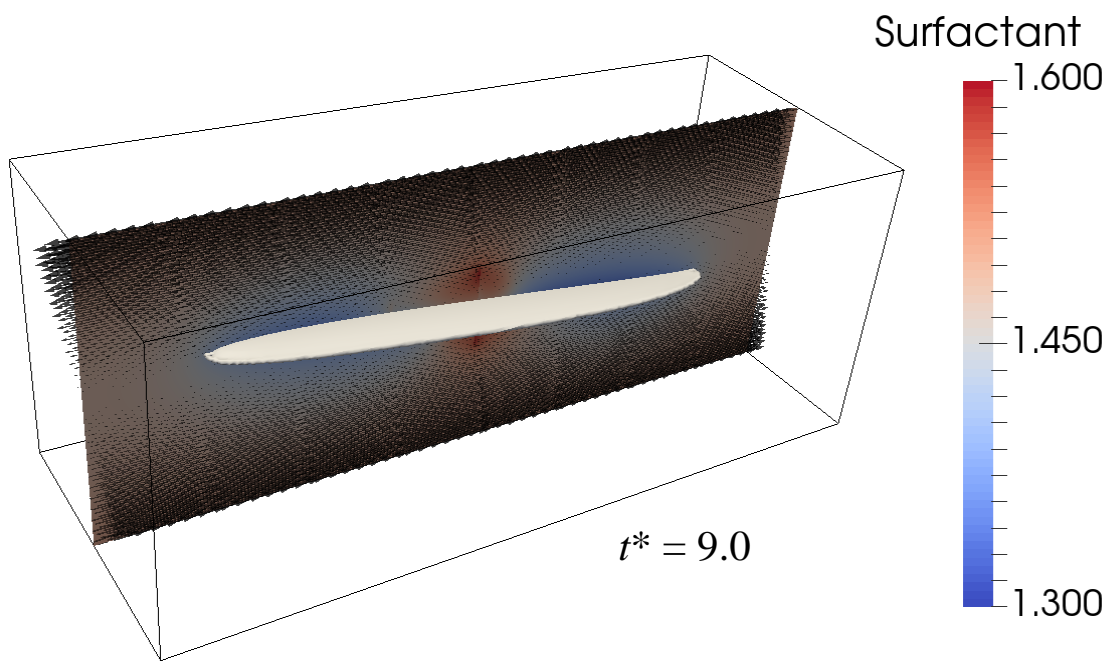

**Fig. 17**

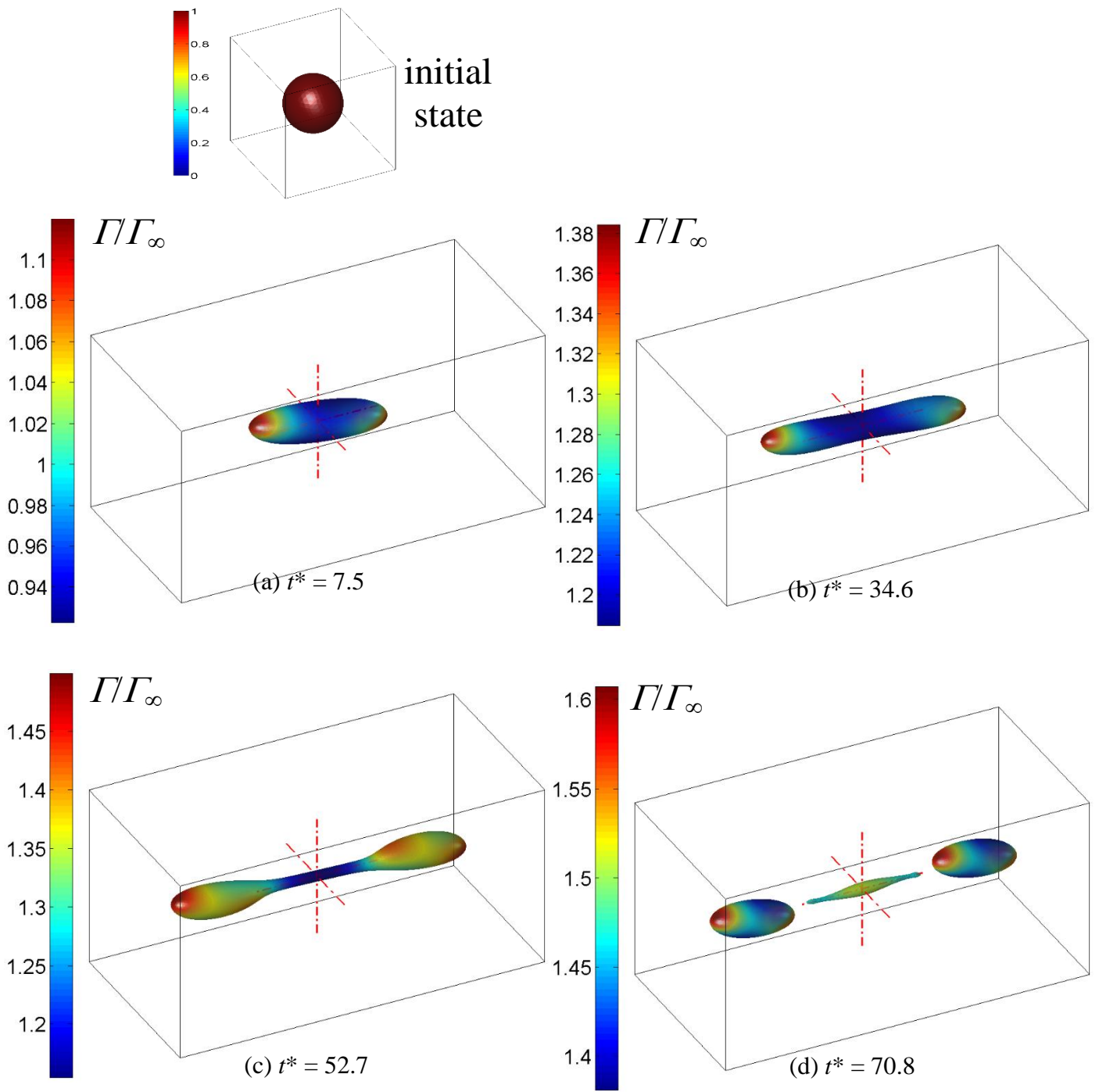

**Fig. 18**

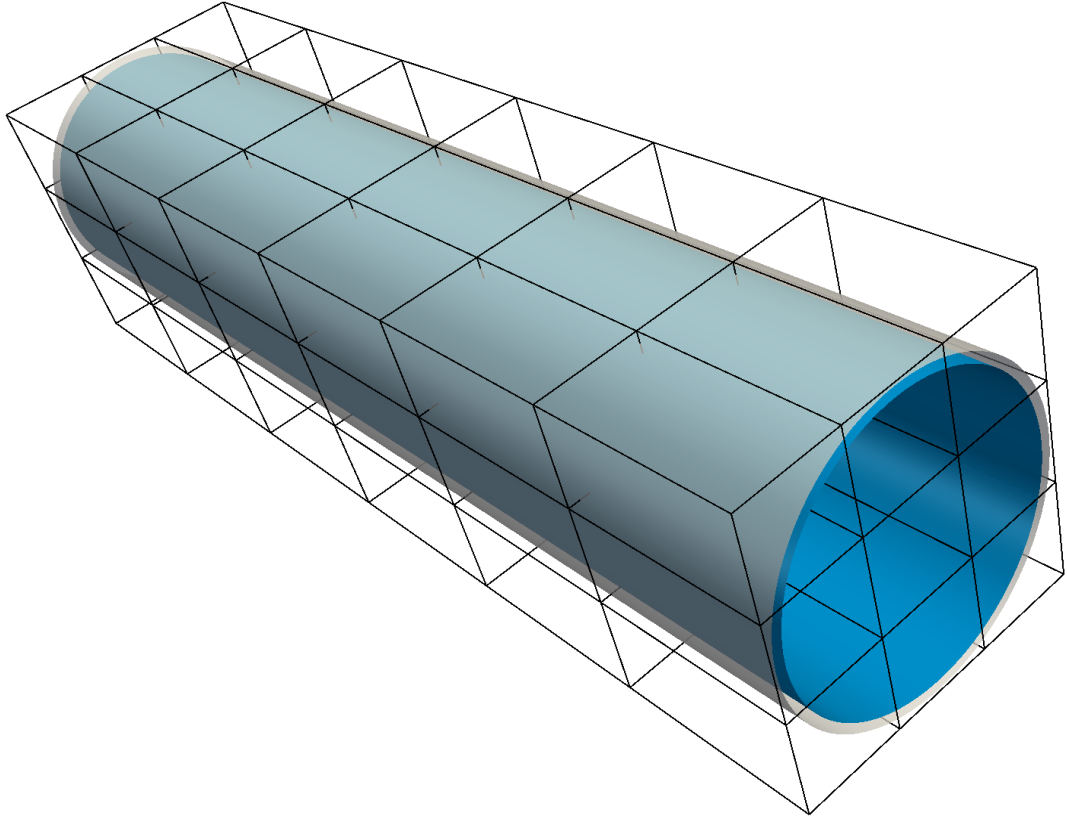

**Fig. 19**

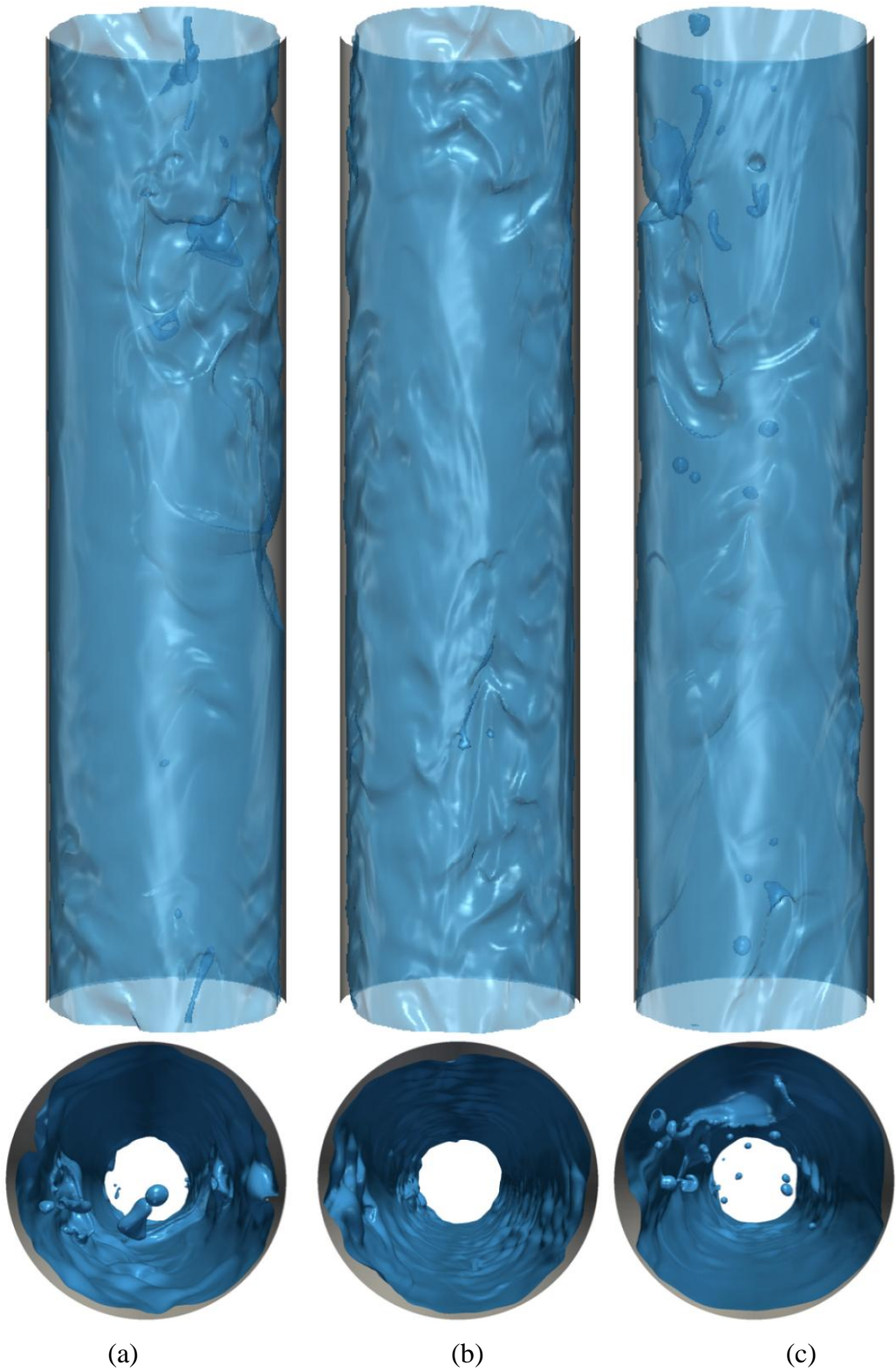

(a)            (b)            (c)

**Fig. 20**

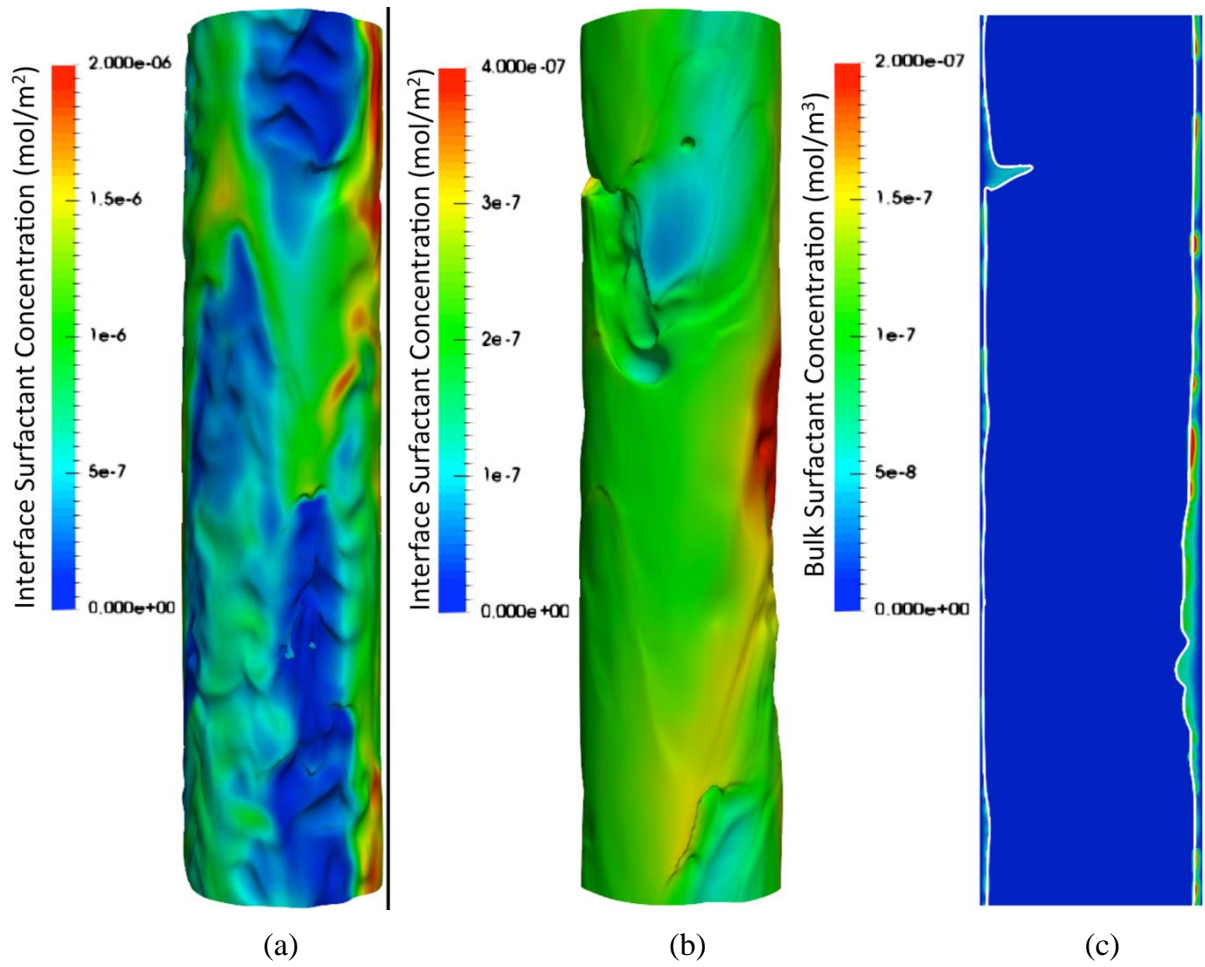

(a)　　　　　　　(b)　　　　　　　(c)

**Fig. 21**

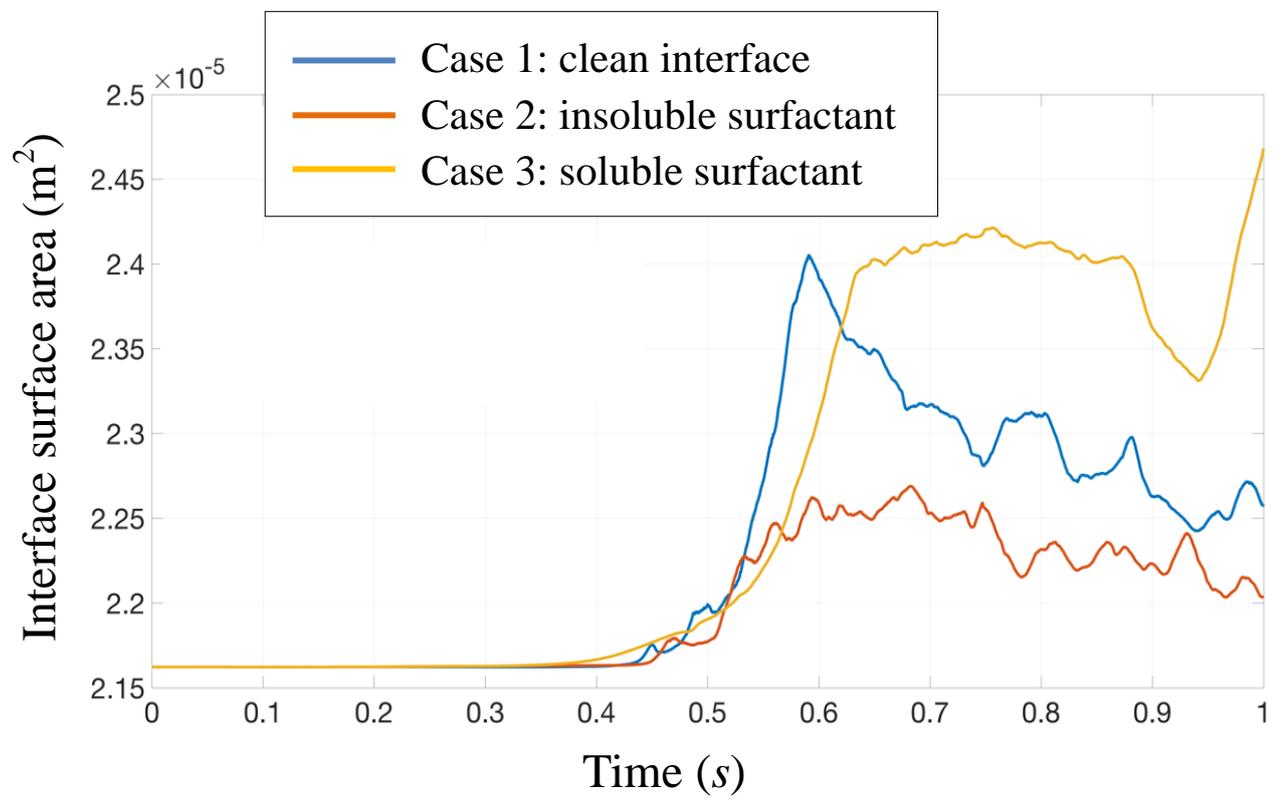

**Fig. 22**